\documentclass[12pt]{iopart}

\usepackage{graphicx}

  \usepackage[utf8]{inputenc}
\usepackage{color}
\usepackage{amssymb}
  \usepackage{amsthm}
  \usepackage{graphicx}
\newsavebox{\astrutbox}
\sbox{\astrutbox}{\rule[-5pt]{0pt}{20pt}}

\newcommand\thalf{\ensuremath{{\textstyle\frac{1}{2}}}}

\begin{document}

\title[Pinch dynamics in a low-$\beta$ plasma]{Pinch dynamics in a low-$\beta$ plasma}

\author{H.\,K.\,Moffatt$^1$\footnote{Corresponding 
author: hkm2@cam.ac.uk} and K.\,Mizerski$^2$\footnote{Corresponding author: krzysztof.mizerski@gmail.com}}

\address{$^1$DAMTP Centre for Mathematical Sciences, University of Cambridge,
Wilberforce Road, Cambridge, CB3 0WA, UK}
\address{$^2$Department of Magnetism, Institute of Geophysics, Polish Academy of Sciences, Ksiecia Janusza 64, 01-452 Warsaw, Poland}

\ead{hkm2@cam.ac.uk}

\begin{abstract}
The relaxation of a helical magnetic field ${\bf B}({\bf x}, t)$ in a  high-conductivity plasma contained in the annulus between two perfectly conducting coaxial cylinders is considered.  The plasma is of low density and its pressure is negligible compared with the magnetic pressure; the  flow of the plasma is driven by the Lorentz force and and energy is dissipated primarily by the viscosity of the medium.  The axial and toroidal fluxes of magnetic field are conserved in the perfect-conductivity limit, as is the mass per unit axial length.  The magnetic field relaxes during a rapid initial stage to a force-free state, and then decays slowly, due to the effect of weak resistivity $\eta$, while constrained to remain approximately force-free.   Interest centres on whether the relaxed field may attain a Taylor state; but under the assumed conditions with axial and toroidal flux conserved inside every cylindrical Lagrangian surface, this is not possible.  The effect of an additional $\alpha$-effect associated with instabilities and turbulence in the plasma is therefore investigated in exploratory manner.  An assumed pseudo-scalar form  of $\alpha$ proportional to $q\,\eta\, ({\bf j}\cdot {\bf B})$  is adopted, where $ {\bf j}=\nabla\times {\bf B}$ and $q$ is an $\mathcal{O}(1)$ dimensionless parameter.  It is shown that, when $q$ is less that a critical value $q_c$, the evolution remains smooth and similar to that for $q=0$; but that if $q>q_c$, 
negative-diffusivity effects  act on the axial component of $\bf B$, generating high-frequency rapidly damped oscillations and an associated transitory appearance of reversed axial field.   However, the scalar quantity $\gamma={\bf j}\cdot {\bf B}/B^2$ remains highly non-uniform, so that again the field shows no sign of relaxing to a Taylor state for which $\gamma$ would have to be constant. \end{abstract}

\vspace{2pc}

\maketitle

\section{Introduction}
Magnetic relaxation is the process by which a magnetic field in a highly conducting fluid seeks a minimum energy state subject to pertinent topological constraints  (Moffatt 1985). In the perfect conductivity limit, the lines of force (`$\bf B$-lines') are frozen in the fluid, a topological constraint represented by the family of magnetic helicity invariants
\begin{equation}\label{helicity}
{\cal H}_V=\int_V {\bf A}\cdot {\bf B}\,{\textnormal d}V,
\end{equation}
where ${\bf B}=\nabla\times {\bf A}$, and $V$ is any Lagrangian volume on whose surface $\partial V$ (with unit normal $\bf n$)\, ${\bf n}\cdot {\bf B}=0$. These invariants represent the conserved degree of linkage of $\bf B$-lines within $V$. Of particular importance is the global magnetic helicity $\cal H$, integrated over the whole domain $\cal D$ of fluid. The magnetic energy
\begin{equation}\label{energy}
M(t) =\frac{1}{2}\int_{\cal D} {\bf B}^2\,{\textnormal d}V
\end{equation}
then has a lower bound  (Arnold 1974)
\begin{equation}\label{bound}
M(t) >  \lambda\, |\cal H|,
\end{equation}
where $\lambda>0$ is a constant that depends on the scale and geometry of $\cal D$.

The particular problem addressed in this paper concerns relaxation in a plasma of extremely low-density $\rho$, in which the fluid pressure $p$ is negligible compared with the magnetic pressure $p_M=\thalf \mu_0 {\bf B}^2$, i.e.~$\beta=p/p_M \ll 1$.  Flow of the plasma is then driven solely by the Lorentz force ${\bf j}\times{\bf B}$.  The low density  implies further that  inertia  is negligible compared with the viscous force in the Navier-Stokes equation, the viscosity $\mu$ being essentially independent of $\rho$ in the limit $\rho\rightarrow 0$.    These approximations have been adopted in a  cartesian model  by Bajer \& Moffatt (2013) who treated relaxation of a single-component field, and by Moffatt (2015) who considered the case of a two-component field with non-zero helicity.  Here we shall consider the situation in a cylindrical geometry, for which the Lorentz force includes the `hoop stress' associated with curvature of the $\bf B$-lines. 

\section{Relaxation in a low-$\beta$ plasma}
We consider a two-component helical field in cylindrical polar coordinates $\{r,\theta, z\}$ of the form
\begin{equation}\label{B_field}
{\bf B}=B_0\left(0,b_{\theta}(r,t),b_{z}(r,t)\right).
\end{equation}
The associated current distribution is given by\footnote{For simplicity of notation, we absorb the conventional constant $\mu_0$ in the definition of $\bf j$.}
\begin{equation}\label{current}
{\bf j}=\nabla\times {\bf B}=B_0\left(0,\,- \frac{\partial b_z}{\partial r}, \,\frac{1}{r}\frac{\partial}{\partial r}(r b_{\theta}) \right).
\end{equation}
The first objective is to determine how the relaxation of this field is constrained by the initial magnetic helicity distribution.  The component $b_\theta$ of the field is responsible for the classic `pinch effect'  (Bennet 1934).  In this scenario, it is natural to suppose that the initial $z$-component of field is uniform, and that the initial $\theta$-component is concentrated near the outer cylindrical boundary, such a field then providing  a radial Lorentz force that tends to drive the plasma inwards.  We shall suppose that the resulting radial motion
\begin{equation}\label{u_field}
{\bf u}=\left(u(r,t),0,0 \right)
\end{equation}
is controlled  by viscosity, which, as indicated above, dominates over inertia when the plasma density is sufficiently small.  We further suppose that the fluid pressure is negligible compared with the magnetic pressure, i.e.~this is a `low-$\beta$' plasma.    Of course, the fluid pressure increases in the inner region where the density increases; the effect of this increase can be included without difficulty in the numerical treatment.

In the perfect conductivity limit $\eta=0$, the magnetic field evolves  according to the `frozen-field' equation
\begin{eqnarray}\label{induction0}
\partial{\bf B}/\partial t  = \nabla\times({\bf u}\times{\bf B}) ,\label{induction_vec}
\end{eqnarray}
and, with the neglect of  inertia and pressure gradient, the Navier-Stokes equation degenerates to
\begin{eqnarray}\label{NS0}
{\bf 0}={\bf j}\times{\bf B}+\mu_s \nabla^2 {\bf u} +\left(\frac{1}{3}\mu_s+\mu_b\right) \nabla(\nabla\cdot{\bf u}),
\end{eqnarray}
where $\mu_s$ and $\mu_b$ are the shear and bulk viscosities.
From (\ref{induction0}) and (\ref{NS0}), an equation may easily be derived for the magnetic energy:
\begin{equation}\label{mag_energy}
dM/dt =-\int_{V}\left[\mu_s(\nabla\times{\bf u})^2+\left(\frac{4}{3}\mu_s+\mu_b\right)(\nabla\cdot{\bf u})^2\right] \, {\textnormal d}V.
\end{equation}
 Thus $M(t)$ is monotonic decreasing and bounded below by (\ref{bound}). Equilibrium is attained only when ${\bf u}\equiv 0$, and then from (\ref{NS0}), \,${\bf j}\times{\bf B}=0$, i.e.~the field is `force-free'.  Hence 
\begin{equation}\label{beta}
{\bf j}=\gamma\,{\bf B},
\end{equation}
for some pseudo-scalar function $\gamma(\bf x)$ satisfying $({\bf B}\cdot\nabla)\, \gamma=0$.

For the particular one-dimensional geometry considered in this paper, it is therefore to be expected that, when the magnetic diffusivity $\eta$ is sufficiently weak, the field will  relax rapidly during an initial stage to a  force-free state (with here $\gamma=\gamma(r)$) that has minimum energy compatible with its initial (conserved) topology.  Minimising energy subject to the \emph {single} topological constraint of conserved global helicity yields a force-free field structure with $\gamma=$cst.,~a condition that provides reversed axial field near the outer boundary (Taylor 1974). However, the dynamical process through which such a reversed field may spontaneously appear is not  revealed by the simple process of seeking a minimum-energy state.  Such a reversal cannot in fact appear for so long as the pinching motion is purely radial. However, it seems possible that instabilities of the basic relaxing field may lead to an $\alpha$-effect, which could conceivably achieve reversal.  We shall explore this possibility in \S 8;  first however, we treat  simple radial relaxation neglecting any instabilities that may be present. 

We suppose that the plasma is contained in the cylindrical annulus $\delta\, <r/a<1$, where $0<\delta<1$;  we shall scale all lengths so that, in effect, $a=1$.  The boundaries $r=\delta$ and $r=1$ are assumed to be thin perfectly conducting cylinders, separating the plasma from the internal and external regions; these boundaries can therefore support current sheets with both $z$- and $\theta$-components.  The electric field  is given by ${\bf E}=\eta \,{\bf j}-{\bf u}\times{\bf B} =(0, E_{\theta}, E_z)$, and we suppose that  ${\bf E}=0$ in the internal and external  regions, assumed insulating, i.e.~for $r<\delta$ and  $r>1$.  Since $u=0$ on both boundaries and both tangential components of ${\bf E}$ are continuous across them, it follows that, with ${\bf n}=(1,0,0)$,
\begin{equation}\label{bc1}
\eta\, {\bf n}\times {\bf j} ={\bf n}\times {\bf E}=0 \quad\textnormal {on} \,\,r=\delta\,\,\textnormal {and on}\,\, r=1,
\end{equation}
i.e.~that
\begin{equation}\label{bc2}
\eta\, \partial b_z/\partial r =0,\quad \eta \,\partial (r b_\theta/)\partial r =0, \quad\textnormal {on}\,\, r=\delta\,\,\textnormal {and on}\,\, r=1.
\end{equation}
Some type of boundary-layer behaviour is to be expected  in the limit $\eta\rightarrow 0$.

\section{Field evolution and flux conservation}
When the magnetic diffusivity $\eta$ is nonzero, the field evolution is described by the induction equation,
\begin{equation}\label{induction}
\frac{\partial{\bf B}}{\partial t}=-\nabla\times {\bf E}=\nabla\times({\bf u}\times{\bf B})+\eta \,\nabla^{2} {\bf B}.
\end{equation}
and the fluid density $\rho$ satisfies the mass conservation equation
\begin{equation}\label{mass_conservation}
\frac{\partial \rho}{\partial t}=-\nabla\cdot(\rho {\bf u}).
\end{equation}
These equations may be combined to give
\begin{equation}\label{Lagrangian_induction}
\frac{D}{Dt}\left(\frac{\bf B}{\rho}\right) = \left(\frac{\bf B}{\rho}\cdot\nabla \right){\bf u} +\frac{\eta}{\rho}\nabla^2{\bf B}\,,
\end{equation}
where $D/Dt\equiv(\partial/\partial t+{\bf u}\cdot\nabla)$, the Lagrangian (or `material') derivative. Noting that, from (\ref{B_field}) and (\ref{u_field}),
\begin{equation}\label{bdotgradu}
({\bf B} \cdot\nabla){\bf u} = B_0(0, b_{\theta} u/r,0),
\end{equation}
when $\eta=0, \,$eqn.(\ref{Lagrangian_induction}) gives
\begin{equation}\label{br_and_btheta}
\frac{D}{Dt}\left(\frac{b_z}{\rho}\right) =0\quad \textnormal {and}\quad \frac{D}{Dt}\left(\frac{b_{\theta}}{r \rho}\right) =0\,,
\end{equation}
i.e.~when following any material element of fluid,  $b_z/\rho$ and $(b_{\theta}/ r)/\rho$ are constant.  

In Eulerian form, eqn.(\ref{induction}) has components
\begin{equation}\label{b_theta_eqn}
\frac{\partial b_\theta}{\partial t}= - \frac{\partial}{\partial r}(u b_\theta) + \eta\frac{\partial}{\partial r}\frac{1}{r}\frac{\partial}{\partial r} (r b_\theta) \,,
\end{equation}
and
\begin{equation}\label{b_z_eqn}
\frac{\partial b_z}{\partial t}= - \frac{1}{r}\frac{\partial}{\partial r}(r u b_z) + \eta\,\frac{1}{r}\frac{\partial}{\partial r} r\frac{\partial b_z}{\partial r}\,.
\end{equation}
The flux of $b_z$ between the two cylinders is
\begin{equation}\label{Flux_z}
\Phi_{z}=\int_{\delta}^1 b_{z}(r,t) \,2\pi r\, {\rm d}r,
\end{equation}
and we note, using (\ref{b_z_eqn}) and the conditions $u=0,\, \eta\,\partial b_z/\partial r=0$ on both boundaries,  that
\begin{equation}\label{Phiz}
\frac{1}{2\pi}\frac{ {\rm d}\Phi_{z}}{ {\rm d}t}\!=\!\!\!\int_{\delta}^1\! \frac{\partial b_{z}}{\partial t} r {\rm d}r\! = \!\!\!\int_{\delta}^1
\!\!\frac{\partial}{\partial r}\!\left(\!\!-r u b_z \!+ \!\eta r\frac{\partial b_z}{\partial r}\!\right)\! {\rm d}r\!=\!\left[ \!- r u b_z \!+\!\eta r\frac{\partial b_z}{\partial r}\right]_{\delta}^1\!=0.
\end{equation}
It follows that $\Phi_{z}=$ cst.

Similarly, the flux of $b_{\theta}$ in the $\theta$-direction, per unit axial length between the cylinders, is
\begin{equation}\label{Flux_theta}
\Phi_{\theta}=\int_{\delta}^1 b_{\theta}(r,t) \, {\rm d}r,
\end{equation}
and it follows in the same way from (\ref{b_theta_eqn}), and the conditions $u=0, \,\eta\,\partial (r b_\theta)/\partial r=0$ on both boundaries,  that
$\Phi_{\theta}=$ cst.~also.  The constraints
\begin{equation}\label{Flux_constraints}
\Phi_{z}= \textnormal{cst.},\quad \Phi_{\theta}= \textnormal{cst.},
\end{equation}
provide an important check on the numerical computations that follow (see Fig.~\ref{Fig_Relaxation00}(f)).  The mass per unit axial length $\mathbb{M}$ is of course similarly constant:
\begin{equation}\label{Mass} 
\mathbb{M} = \int_{\delta}^1 \rho(r,t) \,2\pi r\, {\rm d}r =\textnormal{cst.}
\end{equation}
The results (\ref{Flux_constraints})  are clearly compatible with (\ref{br_and_btheta}) when $\eta=0$.

\section{Initial conditions}
\begin{figure}
\centering
\includegraphics[width=.8 \textwidth,  trim=0mm 190mm 0mm 0mm]{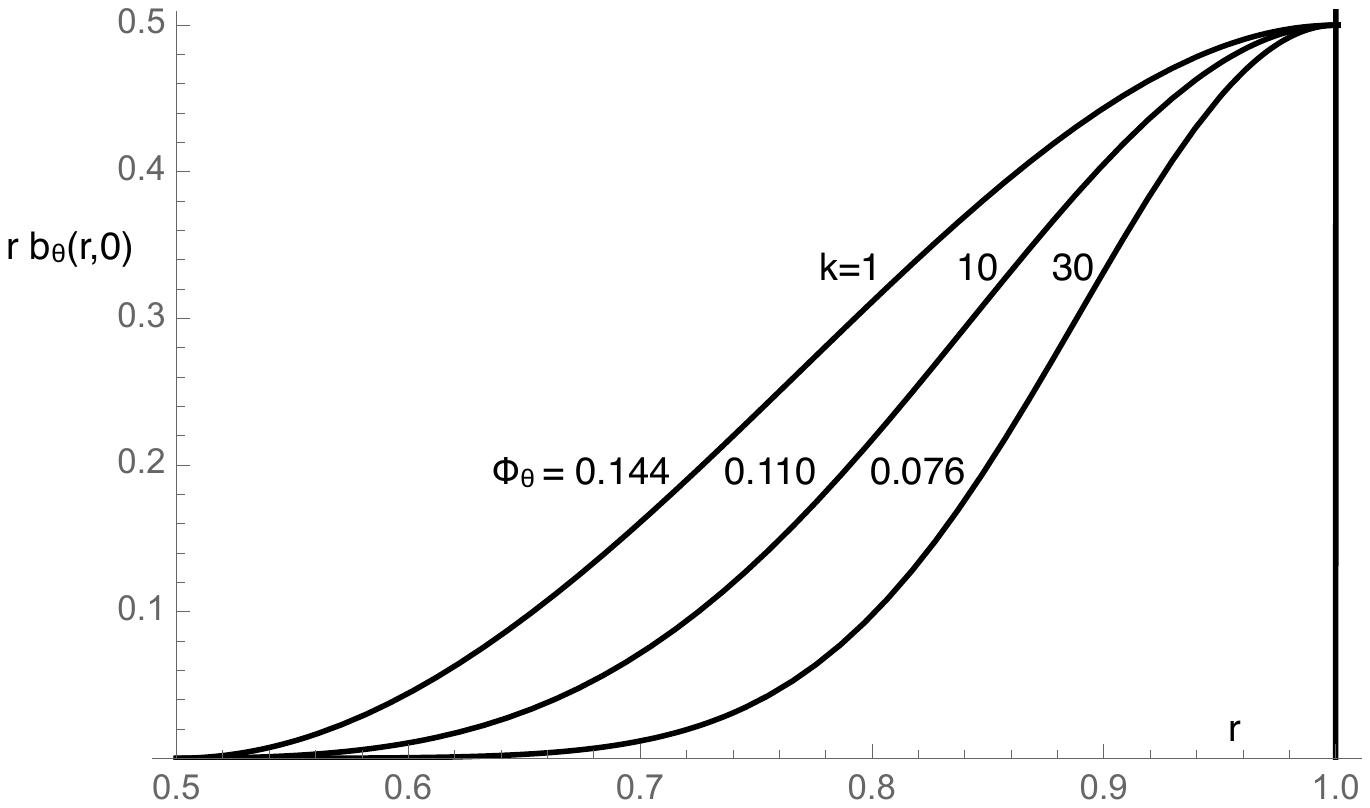}
\caption {Initial profiles of $r b_{\theta}$ for $\delta=0.5, \,c=0.5$, and three values of $k$, with corresponding values of the flux $\Phi_\theta$. } 
\label{Fig_bth0}
\end{figure}

We adopt as initial conditions for the magnetic field
\begin{equation}\label{initial_bz}
 b_z(r,0)= \frac{1}{\pi(1-\delta^2)}\qquad (\textnormal{so}\,\,\Phi_z=1),
\end{equation}
and 
\begin{equation}\label{initial_btheta}
 r\,b_\theta(r,0)=  \frac{c}{3(1-\delta)^{3}} \left[3(r - \delta)^{2}(1-\delta) - 2(r-\delta)^3\right] \textnormal{e}^{-k (1 - r)^{2}}.
\end{equation}
These are chosen to be compatible with  the boundary conditions (\ref{bc2}) for any values of the parameters $\{\delta, c,k\}$, and to satisfy $b_\theta(1,0)=c$.  Fig.~\ref{Fig_bth0} shows the field $r b_\theta(r,0)$ for $\delta=c=0.5$ and for three values $k=1,\,10,\,30$, together with corresponding values of the flux $\Phi_\theta$.   Increasing $k$ leads to increasing concentration of $r b_\theta(r,0)$ near the outer boundary. Rather arbitrarily we choose 
\begin{equation}\label{parameter_values}
\delta=0.5,\quad c=0.5,\quad k=10,
\end{equation}
in the computations that follow.
We further adopt initial conditions for the velocity and density fields,
\begin{equation}\label{initial_u_and_rho}
u(r,0)=0, \qquad \rho(r,0) =\rho_0,
\end{equation}
i.e.~the plasma is initially at rest with uniform density $\rho_0$.

\section{Dynamics of the relaxation process}
The Lorentz force in the cylindrical geometry considered here takes the form
\begin{equation}\label{Lorentz}
{\bf j}\times{\bf B}=B_0^2 \left(-\frac{1}{2}\frac{\partial}{\partial r}(b_{\theta}^2 +b_{z}^2) -\frac{b_{\theta}^2}{r},\,0,\,0\right).
\end{equation}
 The Navier-Stokes equation, including this term, has only a radial component:
\begin{equation}\label{Navier_Stokes}
\rho\!\left(\!\frac{\partial u}{\partial t}+u\frac{\partial u}{\partial r}\! \right)\!=\!-B_0^2\left( \frac{1}{2}\frac{\partial}{\partial r}(b_{\theta}^2 +b_{z}^2) +\frac{b_{\theta}^2}{r}\right)\!+\!\mu\left(\frac{\partial^{2} u}{\partial r^{2}}\!+\!\frac{1}{r}\frac{\partial u}{\partial r}\!-\!\frac{u}{r^2} \right),
\end{equation}
where $\mu=4\mu_s/3+\mu_b$ is an effective viscosity.
As in Moffatt (2015), it is now convenient to introduce dimensionless variables
\begin{eqnarray}\label{dimensionless_variables}
\hat r = r/a,\quad \hat t = t\,B_{0}^2/\mu ,\quad \hat \rho = \rho/\rho_0, \quad \hat u = u\,\mu/B_{0}^2\, a, 
\end{eqnarray} 
and dimensionless parameters
\begin{equation}\label{dimensionless_parameters}
\kappa=\eta\mu/B_0^2 a^2,\qquad \epsilon =\rho_0 B_0^2 a^2/\mu^2;
\end{equation}
we assume that both these parameters are small: $\kappa\!\ll\!1$ (i.e.~small diffusivity); and $\epsilon \ll1$ (i.e.~low density).
With the variables (\ref{dimensionless_variables}), and immediately dropping the hats, eqn.~(\ref{mass_conservation}) is unchanged, while in eqns.~(\ref{b_theta_eqn}) and (\ref{b_z_eqn}), $\eta$ is simply replaced by $\kappa$:
\begin{equation}\label{b_theta_eqn_dim}
\frac{\partial b_\theta}{\partial t}= - \frac{\partial}{\partial r}(u b_\theta) + \kappa\,\frac{\partial}{\partial r}\frac{1}{r}\frac{\partial}{\partial r} (r b_\theta) \,,
\end{equation}
and
\begin{equation}\label{b_z_eqn_dim}
\frac{\partial b_z}{\partial t}= - \frac{1}{r}\frac{\partial}{\partial r}(r u b_z) +\kappa\,\frac{1}{r}\frac{\partial}{\partial r} r\frac{\partial b_z}{\partial r}\,.
\end{equation}
The momentum equation (\ref{Navier_Stokes}) becomes
\begin{equation}\label{NS2}
\epsilon\,\rho\left(\frac{\partial u}{\partial t}+u\frac{\partial u}{\partial r} \right)=- \frac{1}{2}\frac{\partial}{\partial r}(b_{\theta}^2 +b_{z}^2) -\frac{b_{\theta}^2}{r}+\left(\frac{\partial^{2} u}{\partial r^{2}}+\frac{1}{r}\frac{\partial u}{\partial r}-\frac{u}{r^2} \right).
\end{equation}
Here, the term $\thalf\partial(b_{\theta}^2 +b_{z}^2)/\partial r$ is the gradient of magnetic pressure, and the term $-b_{\theta}^2/r$ is the additional `hoop stress' that arises due to curvature of the ${\bf B}$-lines.

\section{Numerical integration}
\begin{figure}
\centering
\includegraphics[width=0.95\textwidth,  trim=-15mm 15mm 0mm 0mm]{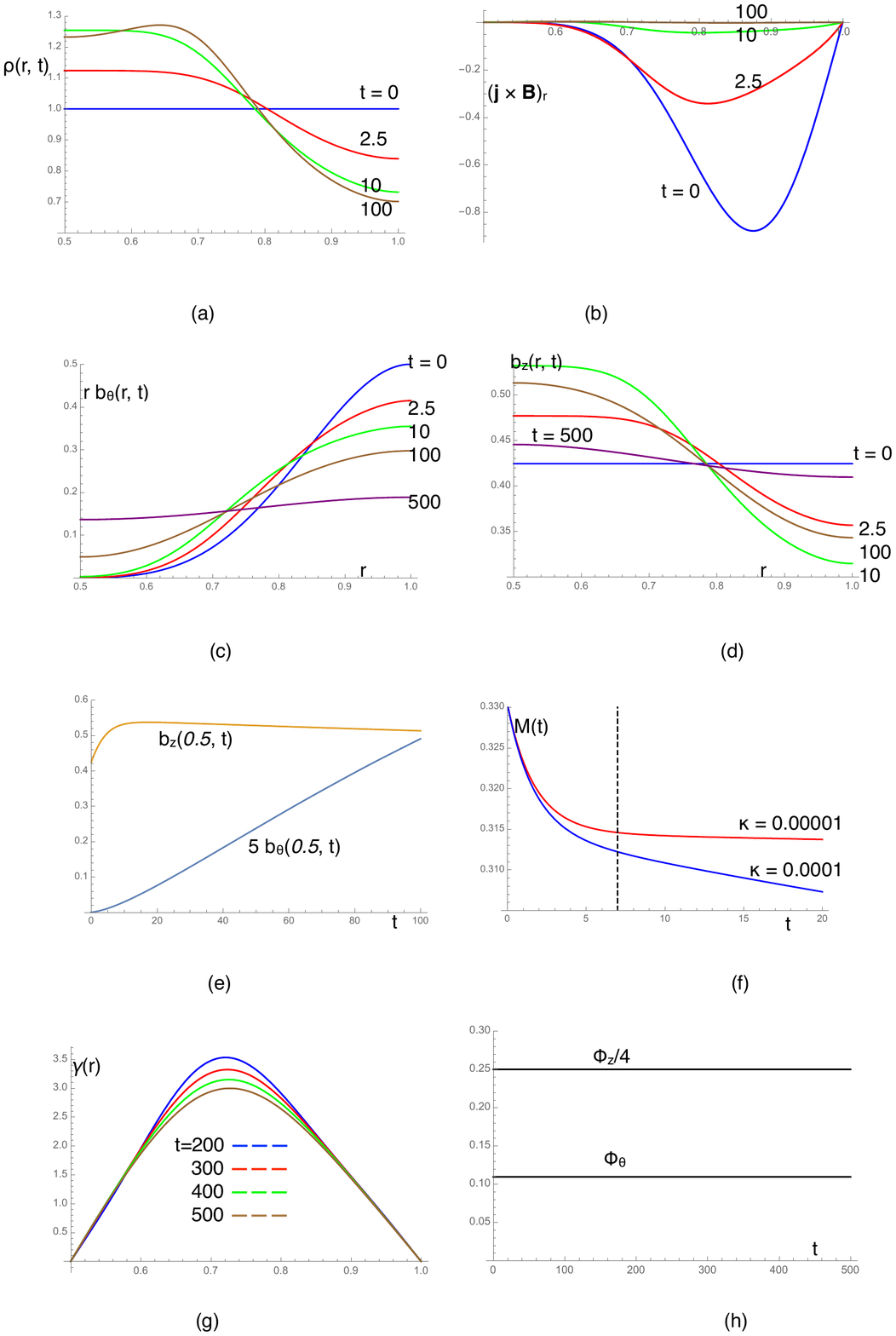}
\caption {Pinch effect with $\delta=c=0.5, k=10, \epsilon=0.01, \kappa=0.0001$; (a) Evolution of density field $\rho(r,t)$; (b) collapse of Lorentz force $({\bf j}\times {\bf B})_r$ during early stage of relaxation; (c) evolution of $r b_{\theta}(r,t)$ and (d) of $b_{z}(r,t)$; (e) the increase of  $5 b_{\theta}(\delta,t)$ and $b_{z}(\delta,t)$ during the initial phase; (f) the decay of magnetic energy; the dashed line separates the early phase from the later diffusive phase of evolution; (g) the function $\gamma(r)=({\bf j}\cdot{\bf B})/{\bf B}^2$ at $t=200,300,400,500$; and (h) the fluxes $\Phi_z$ and $\Phi_\theta$, which remain sensibly constant.}
\label{Fig_Relaxation00}
\end{figure}
We can now proceed to numerical integration of these equations, with boundary conditions as already stated.  When $\kappa\ne 0$, these are
\begin{equation}\label{bc3}
u=0,\quad \partial b_z/\partial r =0,\quad \partial (r b_\theta/)\partial r =0, \quad\textnormal {on}\,\, r=\delta\,\,\textnormal {and on}\,\, r=1,
\end{equation}
and initial conditions as specified in the previous section.  Results obtained with \emph {Mathematica} are summarised in Fig.~\ref{Fig_Relaxation00}(a\,-h), for the particular choice of parameters (\ref{parameter_values}), together with $\epsilon=0.01, \kappa=0.0001$ (the behaviour for this choice is quite typical). The panels of the figure show (a) the inward movement of the density field $\rho(r,t)$ in response  to the negative radial Lorentz force (b), which collapses rapidly to near zero;  (c,d) the corresponding evolution of the magnetic field components;  (e) the rise of both field components at the inner boundary; (f) the decay of magnetic energy;  (g) the pseudo-scalar coefficient $\gamma(r)$ when the nearly force-free state has been established; and finally (h) the fluxes $\Phi_z$ and $\Phi_\theta$ which, as expected, remain constant to within numerical error throughout the whole computational period  $0<t<500$.

The following points are particularly worth noting. First, there are clearly two phases to the evolution: an initial phase (here $0<t\lesssim 7$) during which magnetic diffusion is negligible and the magnetic energy decreases relatively rapidly on the (dimensional) time-scale $\mu/B_{0}^2$; and a  slow diffusive stage $t\gtrsim 7$, during which $b_z$ slowly relaxes back to its initial uniform value (its flux remaining constant); during this phase,  $r b_{\theta}$ also slowly decays to a constant $C\,(=\Phi_{\theta}/\log(1/\delta))$, implying ultimate concentration of axial current  as a current sheet on the inner boundary $r=\delta$.

Second, although not shown in the figure, the energy $\thalf\int b_z^2 \,{\textnormal d}V$ of the $b_z$-field actually \emph {increases} during the initial phase (reaching a maximum at $t\approx 18$), but this increase is more than compensated by the decrease of energy of the $b_{\theta}$-field.  Later, during the slow diffusive stage, both contributions to energy slowly decrease.   Fig.~\ref{Fig_Relaxation00}(f) shows the decay of energy for $\kappa=0.00001$ as well as for $\kappa=0.0001$; as expected, the diffusive effect is less apparent in the former case, but the initial (non-diffusive) stage is quite similar.

Third, as previously noted, when the Lorentz force is effectively zero (i.e.~during the diffusive phase), ${\bf j}=\gamma(r,t){\bf B}$; the coefficient $\gamma$ is then given by 
\begin{equation}\label{gamma}
\gamma=({\bf j}\cdot{\bf B})/{\bf B}^2.
\end{equation}
This coefficient, shown in Fig.~\ref{Fig_Relaxation00}(g), is far from uniform in $r$,  so this is certainly not a Taylor state (for which $\gamma$ would necessarily be uniform). The weak time-dependence of $\gamma$ results from slow continuing  evolution during the diffusive stage. This stage is interesting because the field components continuously adjust themselves in such a way that the force-free condition is maintained; in other words, this is not a pure diffusive process, but one that is still constrained to remain nearly force-free through the dynamics encapsulated in eqn.~(\ref{NS2}).

\section{Limiting behaviour as $\eta\rightarrow 0$}
\begin{figure}
\centering
%[width= \textwidth,  trim=0mm 140mm 0mm 0mm]
\includegraphics[scale=0.123]{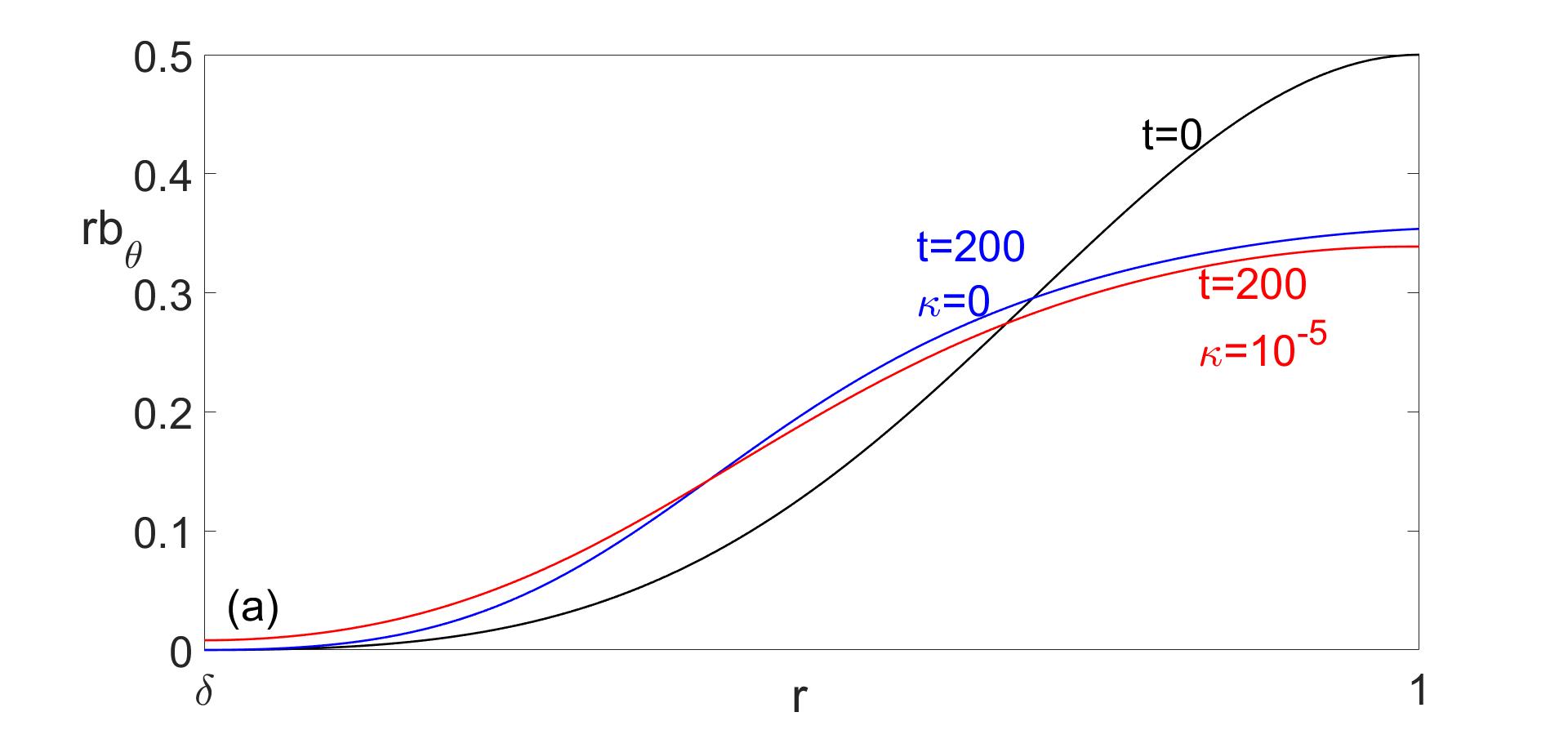}\includegraphics[scale=0.123]{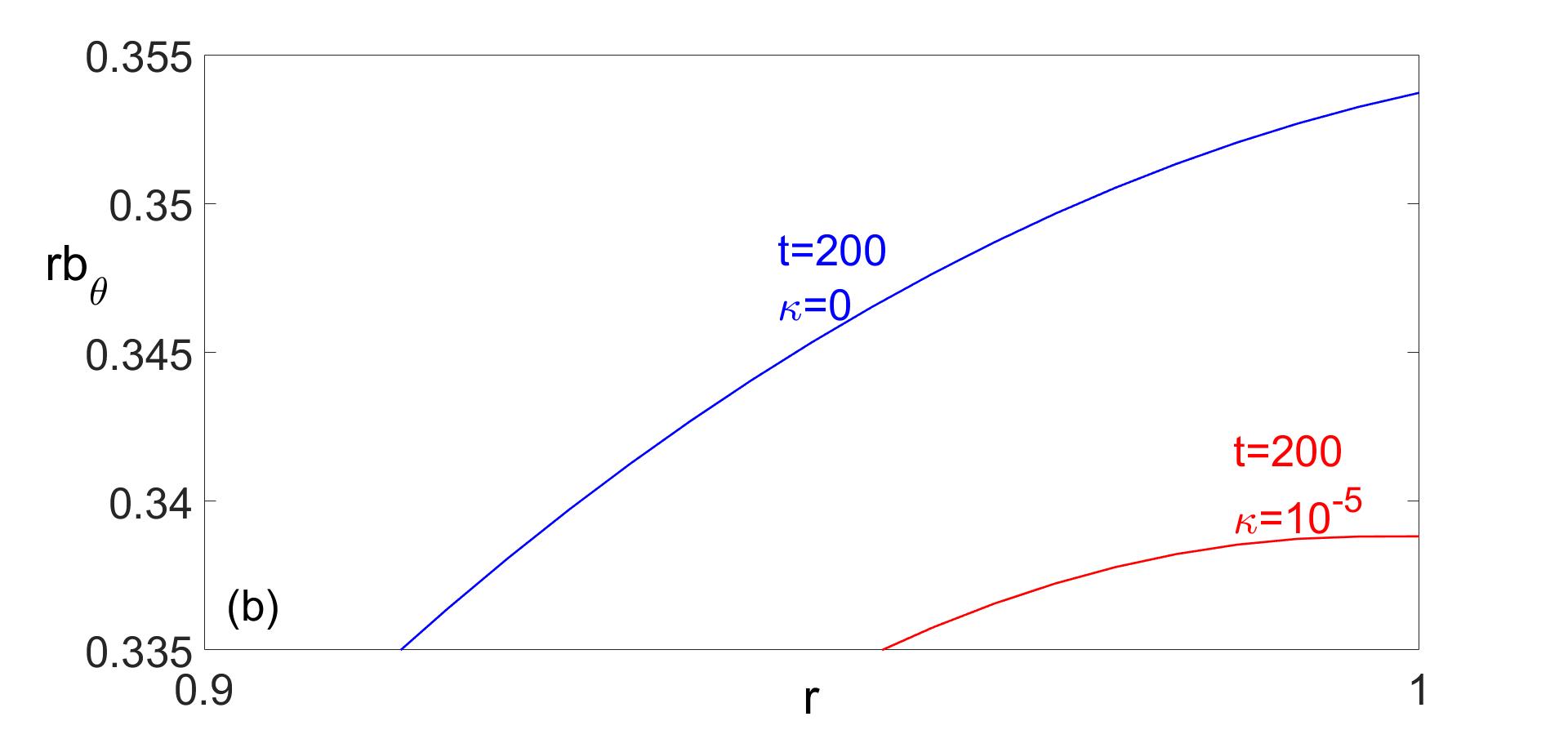}
\includegraphics[scale=0.123]{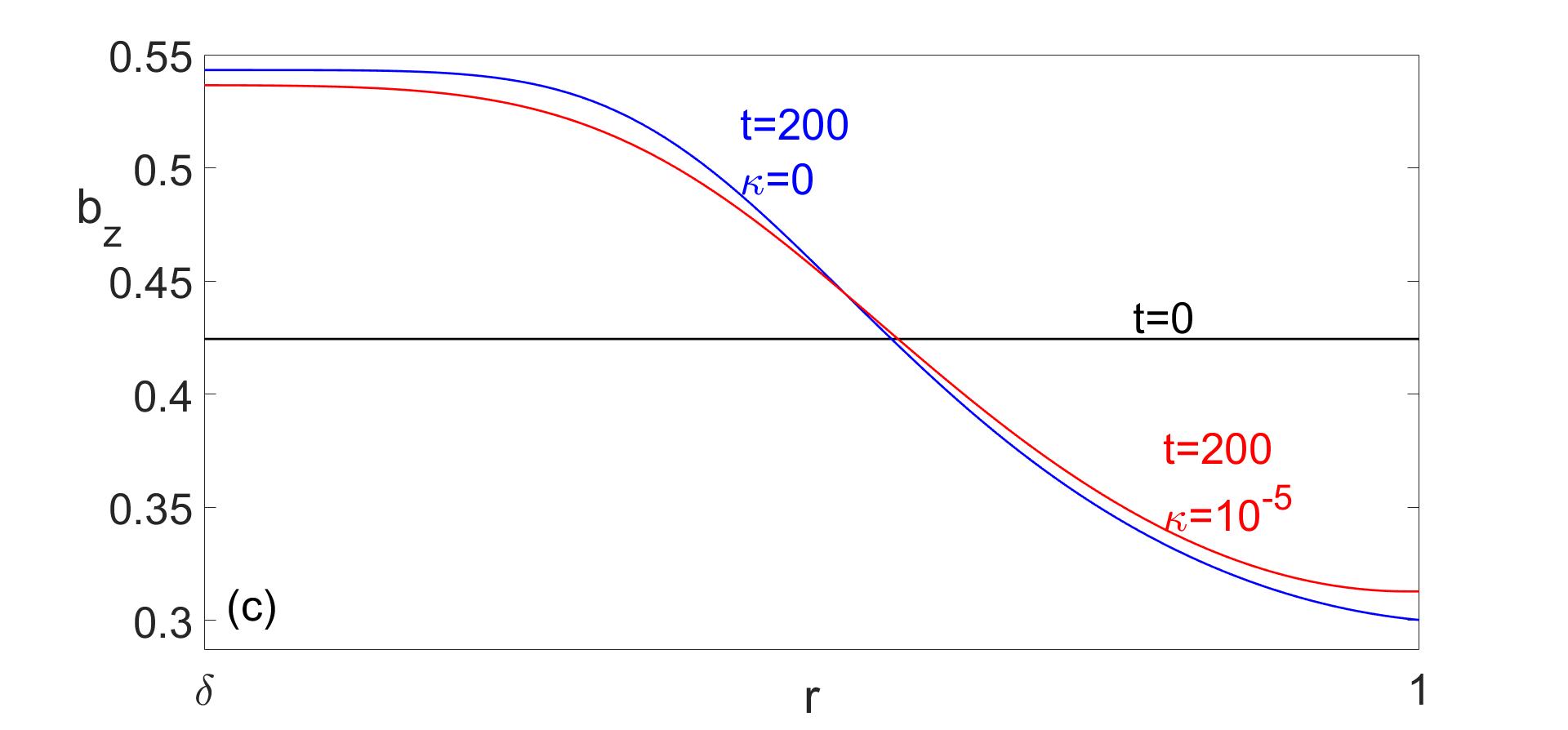}\includegraphics[scale=0.123]{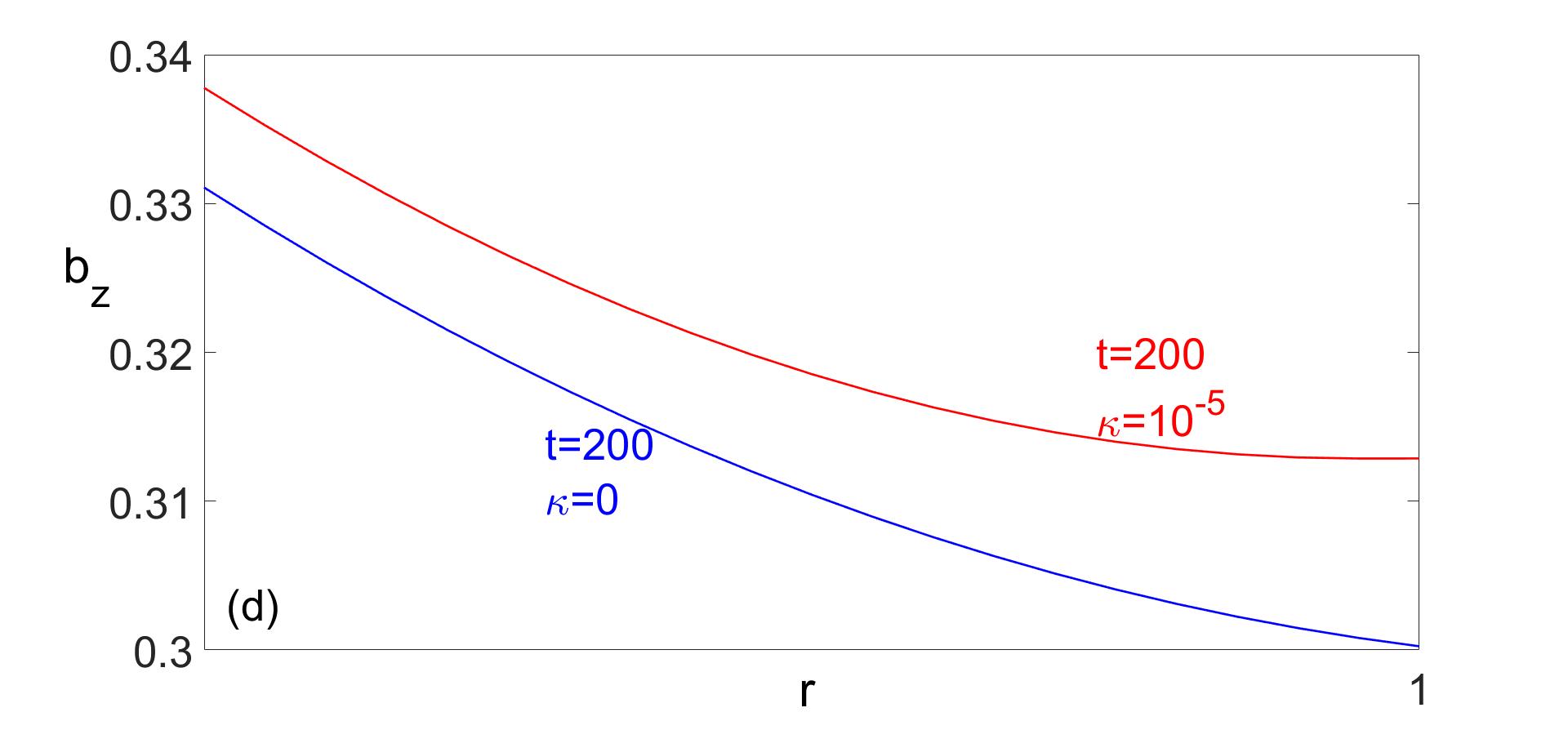}
\caption {Comparison of situation when $\kappa=0$ and $\kappa=0.00001$: (a) $r b_{\theta}(r, t)$ at $t=200$ for $\kappa=0$ (blue) and $\kappa=0.00001$ (red); (b) expanded view of the same curves in the region $0.9<r<1$; (c) $b_z(r, t)$ at $t=200$ for $\kappa=0$ (blue) and $\kappa=0.00001$ (red); (d) expanded view of the $b_z$-curves in the region $0.9<r<1$;  } 
\label{Fig_Limit_behaviour}
\end{figure}
In the limit $\eta =0$, i.e.~$\kappa=0$, we drop the diffusion terms in (\ref{b_theta_eqn_dim}) and (\ref{b_z_eqn_dim}), and  only the boundary condition $u=0$ of (\ref{bc3}) survives,  (\ref{bc2}) being then automatically satisfied.  The two fluxes $\Phi_z$ and $\Phi_\theta$ are still conserved, within numerical error, as in Fig.~\ref{Fig_Relaxation00}. Comparison with the situation when $\kappa=0.00001$ is interesting. The difference is in fact very slight  up to about $t=100$, but becomes visible, particularly near the boundary $r=1$ by $t=200$, as shown in Fig.~\ref{Fig_Limit_behaviour}(a,b).  The expanded views in Figs.~\ref{Fig_Limit_behaviour}(b,d) show that indeed when $\kappa=0$ the boundary condition $\partial (r b_{\theta})/\partial r =0$ is \emph {not} maintained at $r=1$;   when $\kappa=0.00001$ a weak boundary layer is required, within which the solution adapts to this boundary condition.  
 
As time advances, the curve for $\kappa=0.00001$ continues to evolve due to weak diffusion towards the situation $r b_{\theta}=C$, whereas the curve for $\kappa=0$ remains static. In this case of $\kappa=0$, there is only the initial phase of non-diffusive relaxation to a force-free minimum-energy field, which then remains essentially static.
 
\section{Inclusion of an $\alpha$-effect}
It is obvious that purely radial flow cannot lead to local reversal of an axial field that is initially uniform, even in conjunction with diffusion.  However, the situation considered by Taylor (1974) in the context of the reversed-field pinch was that of a fully turbulent plasma in which more complex processes must be present.   Such turbulence presumably results from persistent instability of the evolving mean field $\bf B$, which has  non-zero helicity  represented both by the integrals (\ref{helicity}) and also by the non-zero pseudo-scalar ${\bf j}\cdot {\bf b}$. Such instabilities must inherit the helicity of this mean field, and may be expected to provide an $\alpha$-effect, such that the (non-dimensionalised) mean electric field becomes
\begin{equation}
{\bf E} = \alpha \,{\bf b} - {\bf u}\times {\bf b} +\kappa \,{\bf j},
\end{equation}
where $\bf u$ still represents the mean radial flow, and $\kappa$ now includes turbulent as well as molecular diffusivity.   A stability analysis following the approach of Furth, Killeen \& Rosenbluth (1963) has been carried out by Mizerski (2017), and a resulting anisotropic $\alpha$-effect deduced.  Here, we adopt the simpler isotropic prescription
\begin{equation}\label{alpha}
\alpha(r,t)=  q\, \kappa\,{\bf j}\cdot {\bf b},
\end{equation}
where $q$ is a pure scalar constant (positive or negative); the factor $\kappa$ is included here in recognition of the diffusive origin of the $\alpha$-effect.  The choice (\ref{alpha}) of course ensures that $\alpha$ has the same pseudo-scalar character as ${\bf j}\cdot {\bf b}$.

With this prescription for $\alpha$, the boundary condition ${\bf n}\times {\bf E}=0$ becomes
\begin{equation}\label{bc_alpha}
\kappa \,{\bf n}\times\left( q \,{(\bf j}\cdot {\bf b}){\bf b}  + {\bf j}\right)=0 \quad \textnormal{on}  \,\,r=\delta,\,1,
\end{equation}
and, in the cylindrical geometry,  this is still obviously satisfied by $j_{\theta}=j_z=0$ on $r=\delta, 1$.
The only modification required is therefore the inclusion of a term
\begin{equation}
\nabla\times(\alpha\, {\bf b})=\left(0,\, -\frac{\partial (\alpha \,b_z)}{\partial r}, \,\frac{1}{r}\frac{\partial}{\partial r}(\alpha \,r b_{\theta})\right)
\end{equation}
in the induction equation. In exploratory vein, we include only the z-component\footnote{Justification is suggested by study of the interaction of tearing modes proportional to $\exp i(k_{z} z +m\theta)$ (Mizerski 2017,  particularly eqn.~(44)). It is found that the   $\theta$- and $z$- components of  the curl of the resulting mean emf differ by a factor proportional to $k_z/m$, and that, summing over all Fourier modes, one of these components must vanish.}, thus replacing (\ref{b_z_eqn_dim}) by
\begin{equation}\label{b_z_eqn_modified}
\frac{\partial b_z}{\partial t}= - \frac{1}{r}\frac{\partial}{\partial r}(r u b_z) +\kappa\,\frac{1}{r}\frac{\partial}{\partial r} r\frac{\partial b_z}{\partial r}+q\,\kappa\,\frac{1}{r}\frac{\partial}{\partial r}\left[({\bf j}\cdot{\bf b}) \,r b_{\theta} \right],
\end{equation}
while leaving (\ref{b_theta_eqn_dim}) unaltered. From (\ref{B_field}) and (\ref{current}), we then have
\begin{equation}\label{b_z_eqn_final}
\frac{\partial b_z}{\partial t}= - \frac{1}{r}\frac{\partial}{\partial r}(r u b_z)\! +\!\kappa\,\frac{1}{r}\frac{\partial}{\partial r} r\frac{\partial b_z}{\partial r}\!+\!q\,\kappa\,\frac{1}{r}\frac{\partial}{\partial r}\left[b_z b_{\theta}\frac{\partial}{\partial r}\left( r b_{\theta}\right) \!-\! r b_{\theta}^2 \frac{\partial b_z}{\partial r} \right],
\end{equation}
\begin{equation}\label{b_theta_eqn_final}
\frac{\partial b_\theta}{\partial t}= - \frac{\partial}{\partial r}(u b_\theta) + \kappa\,\frac{\partial}{\partial r}\frac{1}{r}\frac{\partial}{\partial r} (r b_\theta),
\end{equation}
\begin{equation}\label{NS_eqn_final}
\epsilon\rho\!\left(\frac{\partial u}{\partial t}+u\frac{\partial u}{\partial r} \right)=- \frac{1}{2}\frac{\partial}{\partial r}(b_{\theta}^2 +b_{z}^2) -\frac{b_{\theta}^2}{r}+\left(\frac{\partial^{2} u}{\partial r^{2}}+\frac{1}{r}\frac{\partial u}{\partial r}-\frac{u}{r^2} \right),
\end{equation}
\begin{equation}\label{rho_eqn_final}
\frac{\partial \rho}{\partial t}= - \frac{1}{r} \frac{\partial}{\partial r}(r \rho u).
\end{equation}
where, for convenience, we include  the remaining unaltered evolution equations.
%%%%%%%%%%%%%%%%%%%%%%%%%%%%%%%%%%%%%%%%%%%%%%%%%%%%%%%%%%%%%%%%%%%%%%%
%%%%%%%%%%%%%%%%%%%%%%%%%%%%%%%%%%%%%%%%%%%%%%%%%%%%%%%%%%%%%%%%%%%%%%%
%%%%%%%%%%%%%%%%%%%%%%%%%%%%%%%%%%%%%%%%%%%%%%%%%%%%%%%%%%%%%%%%%%%%%%%
%%%%%%%%%%%%%%%%%%%%%%%%%%%%%%%%%%%%%%%%%%%%%%%%%%%%%%%%%%%%%%%%%%%%%%%
%%%%%%%%%%%%%%%%%%%%%%%%%%%%%%%%%%%%%%%%%%%%%%%%%%%%%%%%%%%%%%%%%%%%%%%

\subsection{Field reversal and negative diffusivity}\label{Frnd}
Consider first whether and under what circumstances reversals of the axial field component $b_z$ may occur. Suppose that this does happen, and that the reversal is initiated  at a critical time $t^*$; then  $b_z(r,t^*)$ must equal zero for some $r=r^*$ where $r^*$ may be an internal point or an end-point of the closed interval $[{\delta,1}]$. Moreover, for $t<t^*$, $b_z(r,t)>0 $ for all $r\in[\delta,1]$. Then, at $t=t^*$, we have
\begin{equation}\label{criticality1}
b_z(r, t^*)=0,\,\, \partial b_z(r, t^*)/\partial r =0, \,\,\textnormal{at}\,\, r=r^*,
\end{equation}
and 
\begin{equation}\label{criticality2}
\partial^{2} b_z(r, t^*)/\partial r^{2} >0\, \,\,\textnormal{at}\,\, r=r^*,
\end{equation}
since the curvature is necessarily positive at this point. From eqn.(\ref{b_z_eqn_final}), we then have
\begin{equation}\label{criticality3}
\left[\frac{\partial b_z(r^*,t)]}{\partial t}\right]_{t=t^*}=\kappa\left(1-q\, [b_{\theta}(r^*,t^*)]^2\right)\left[\frac{\partial^{2}b_z(r,t^*)}{\partial r^2}\right]_{r=r^*},\,\,
\end{equation}
all other terms vanishing by virtue of  (\ref{criticality1}). Hence it would appear that $b_{z}(r,t)$ will indeed become negative in a neighbourhood of $r^*$, provided
\begin{equation}\label{criticality4}
q\,[b_{\theta}(r^*,t^*)]^2>1.
\end{equation}
This can therefore  occur only if $q>0$ and $b_{\theta}$ is sufficiently strong; and it is likely to occur first in the region where $|b_{\theta}|$ is maximal. 
More generally, (\ref{b_z_eqn_final}) may be written
\begin{eqnarray}\label{criticality5}
\frac{\partial b_z}{\partial t}= - \frac{1}{r}\frac{\partial}{\partial r}\left(r u b_z\right) &+&\kappa\,\left(1-q\,b_{\theta}^2\right)\frac{1}{r}\frac{\partial}{\partial r} r\frac{\partial b_z}{\partial r}\nonumber\\
 &+&\frac{q\,\kappa}{r}\frac{\partial}{\partial r}\left[b_z b_{\theta}\frac{\partial}{\partial r}\left( r b_{\theta}\right)\right] - q\,\kappa \frac{\partial b_z}{\partial r}\frac{\partial }{\partial r}\left(b_{\theta}^2\right) .
\end{eqnarray}
The second term on the right-hand side has a diffusive character, but with negative diffusivity in any region where $q \,b_{\theta}^2 >1$.  (The remaining terms of the right-hand side involve only $b_z$ and $\partial b_z/\partial r$.)  It follows that $b_z$ can become negative only if this is `triggered' by a  period of negative diffusivity in some  $r$-interval.

We continue to use the parameter values (\ref{parameter_values}).  With these values, $b_{\theta}(r,0)$ is maximal at $r\approx 0.9765$, with maximum value $0.5059$; an interval of negative diffusivity therefore occurs for $q \gtrsim 3.907$.  

\subsection{Results for $q\leq 4$}

\begin{figure}
\centering
%[width= \textwidth,  trim=0mm 140mm 0mm 0mm]
\includegraphics[scale=0.123]{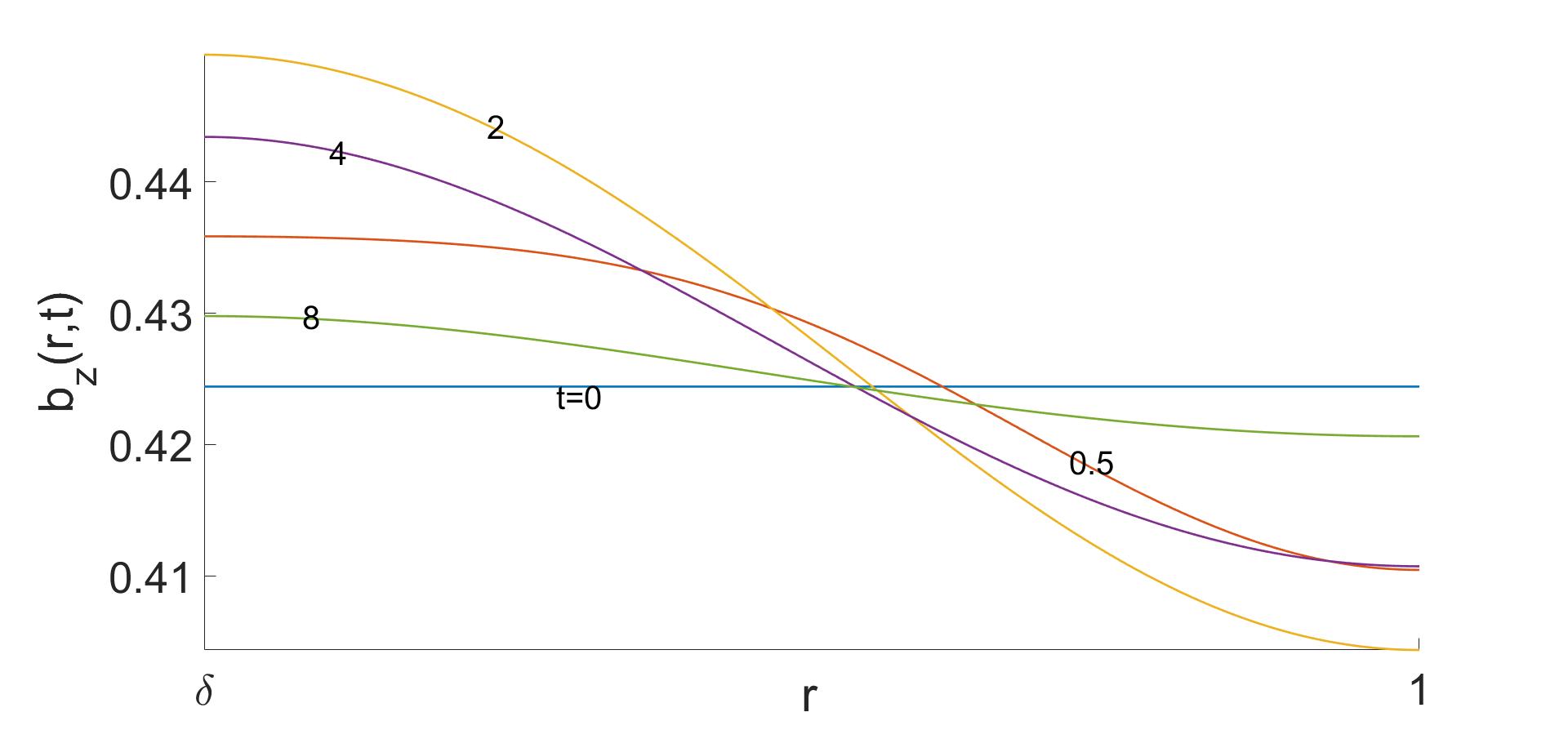}\includegraphics[scale=0.123]{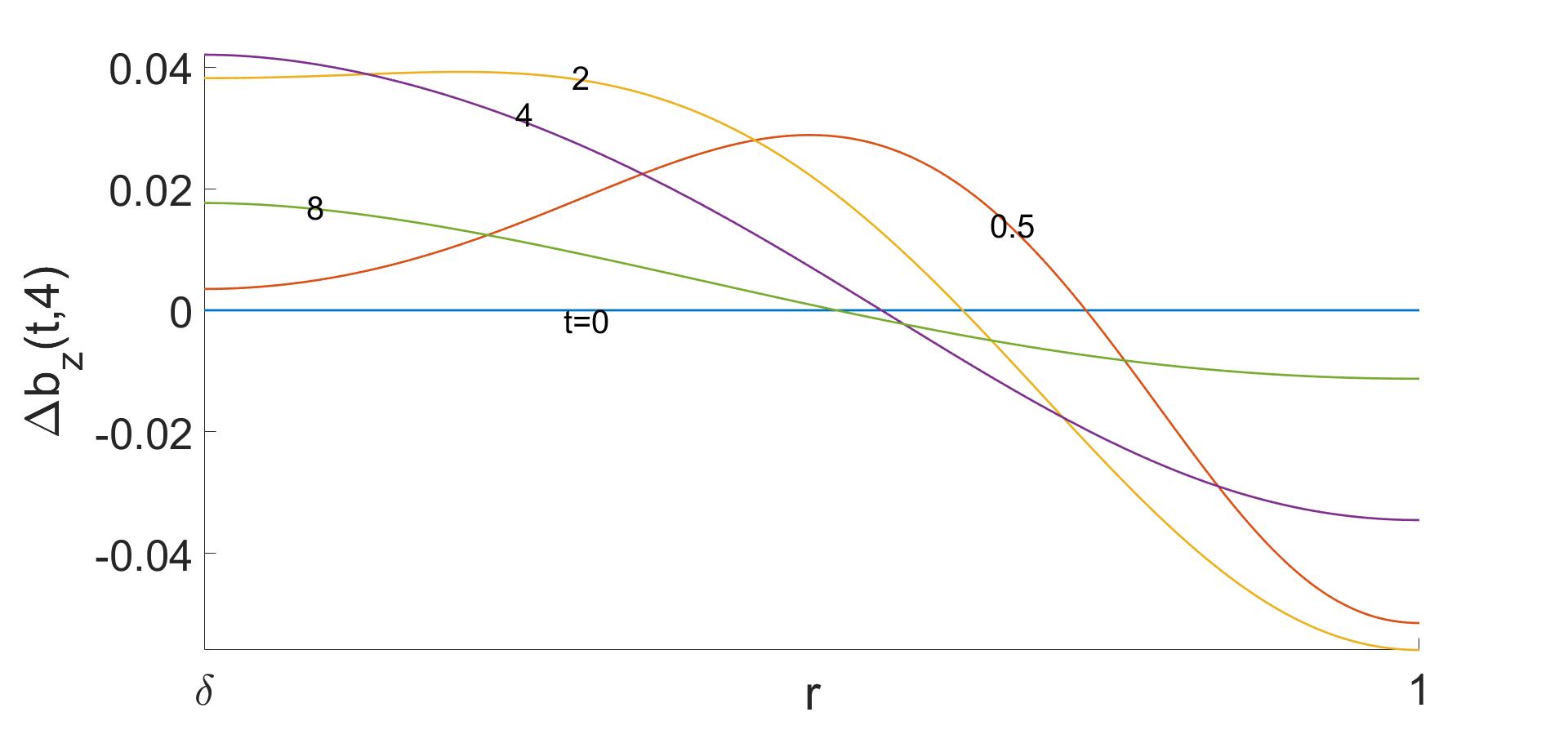}
\includegraphics[scale=0.123]{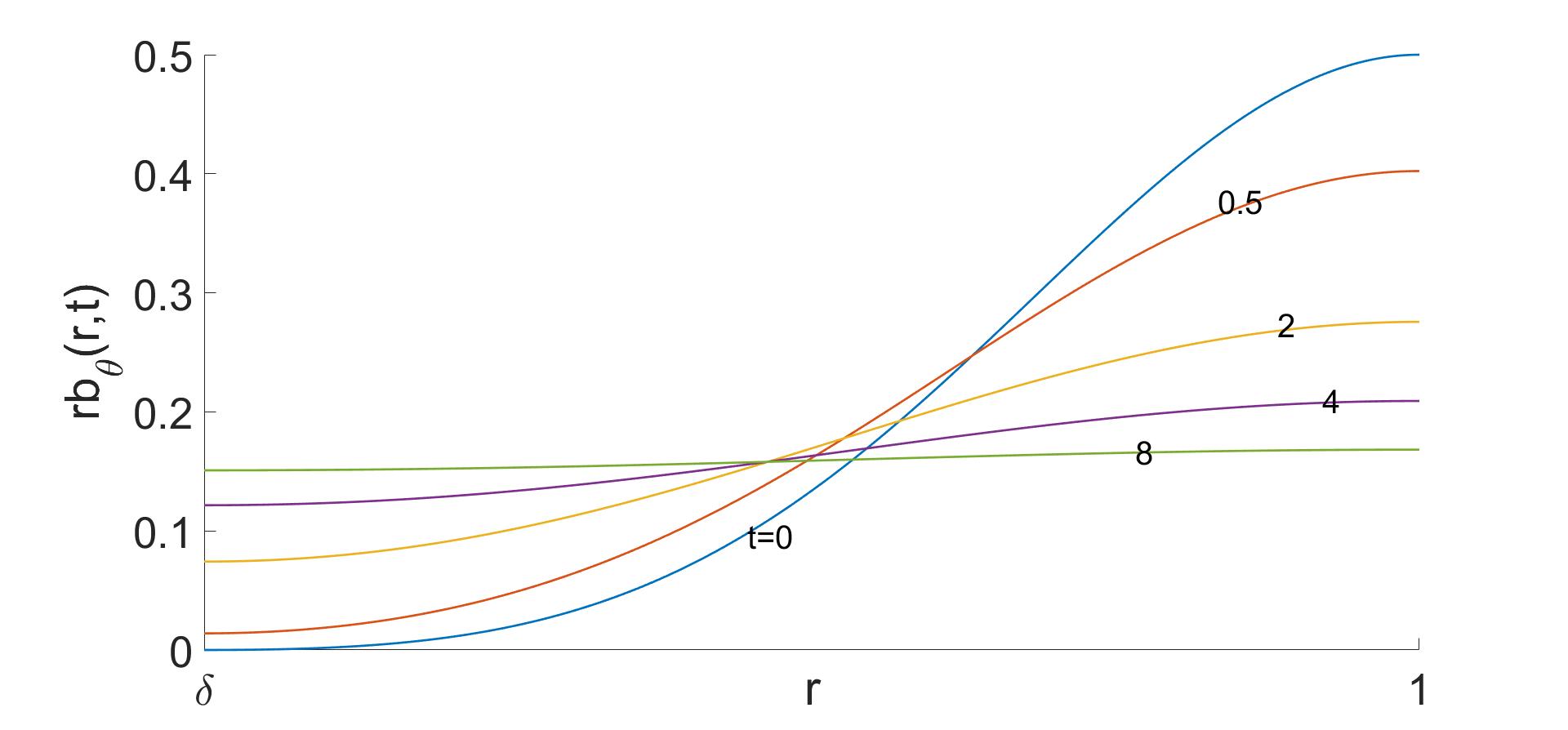}\includegraphics[scale=0.123]{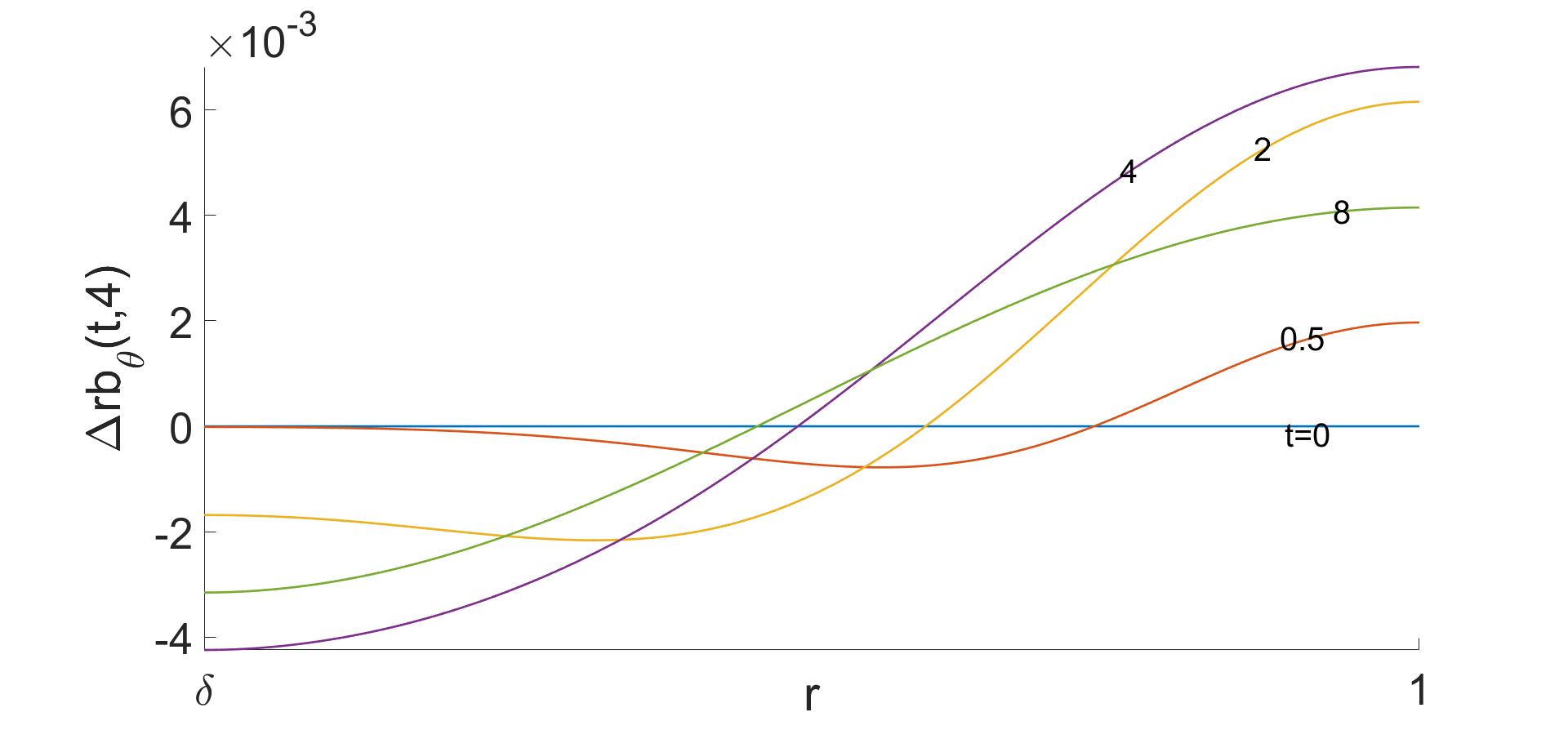}
\includegraphics[scale=0.123]{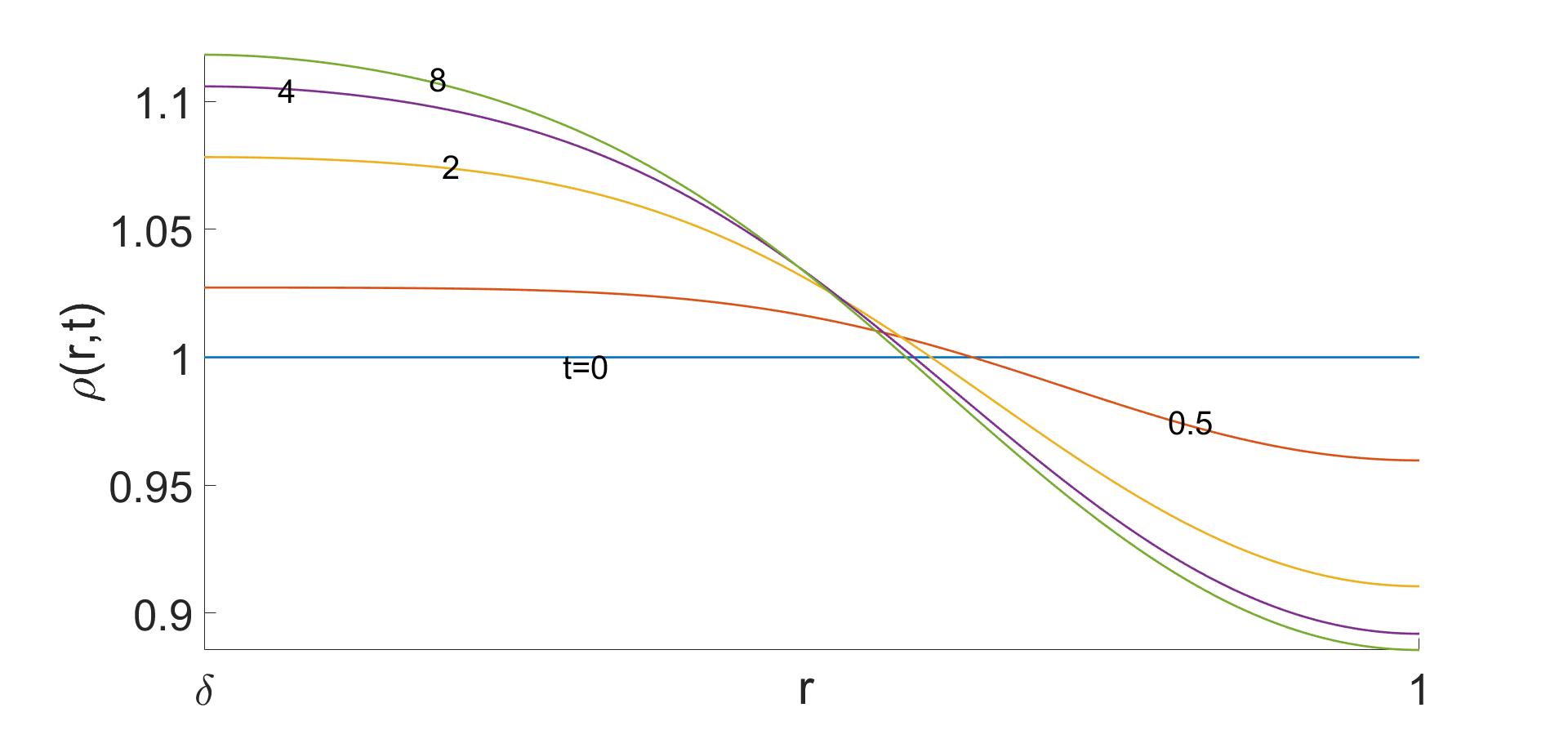}\includegraphics[scale=0.123]{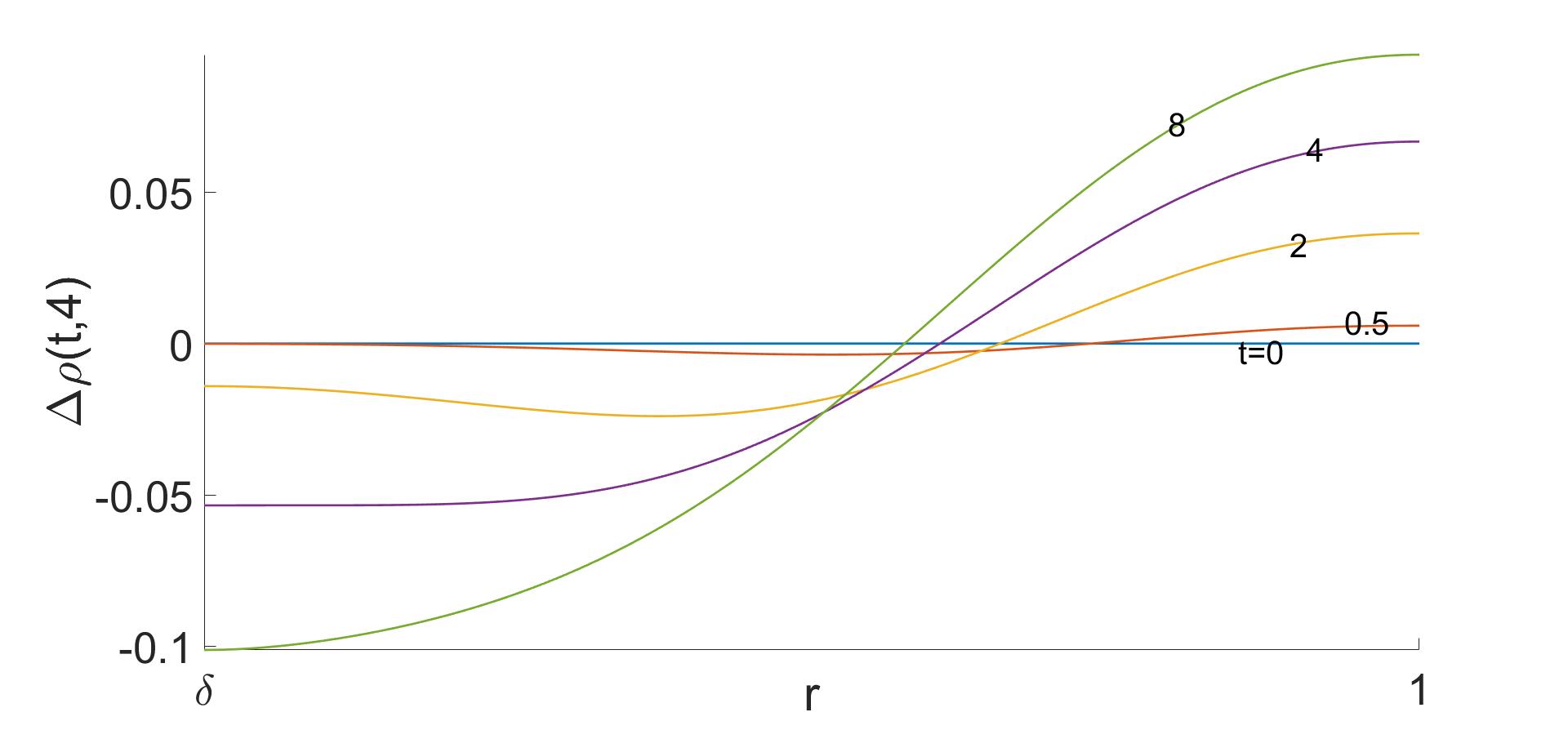}
\includegraphics[scale=0.123]{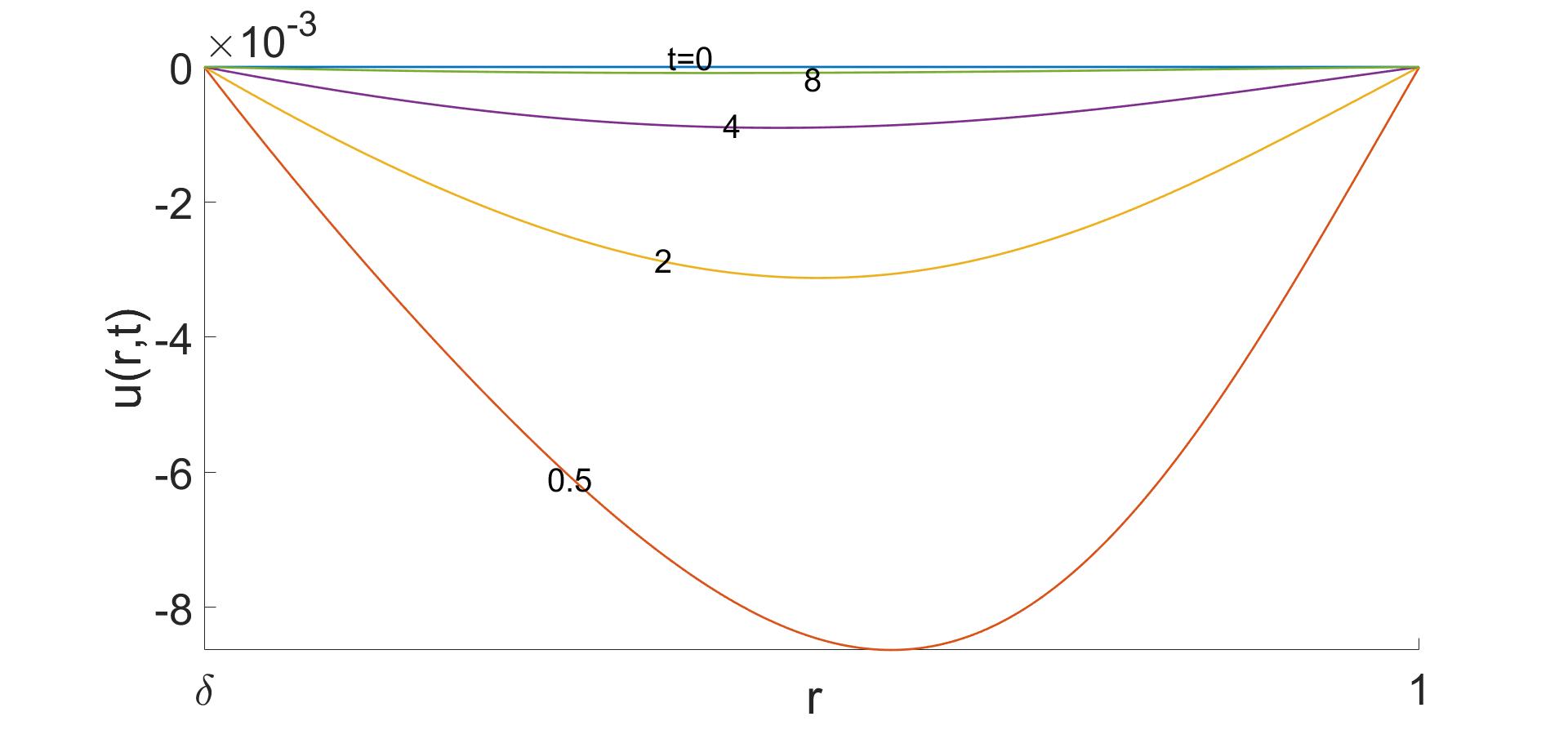}\includegraphics[scale=0.123]{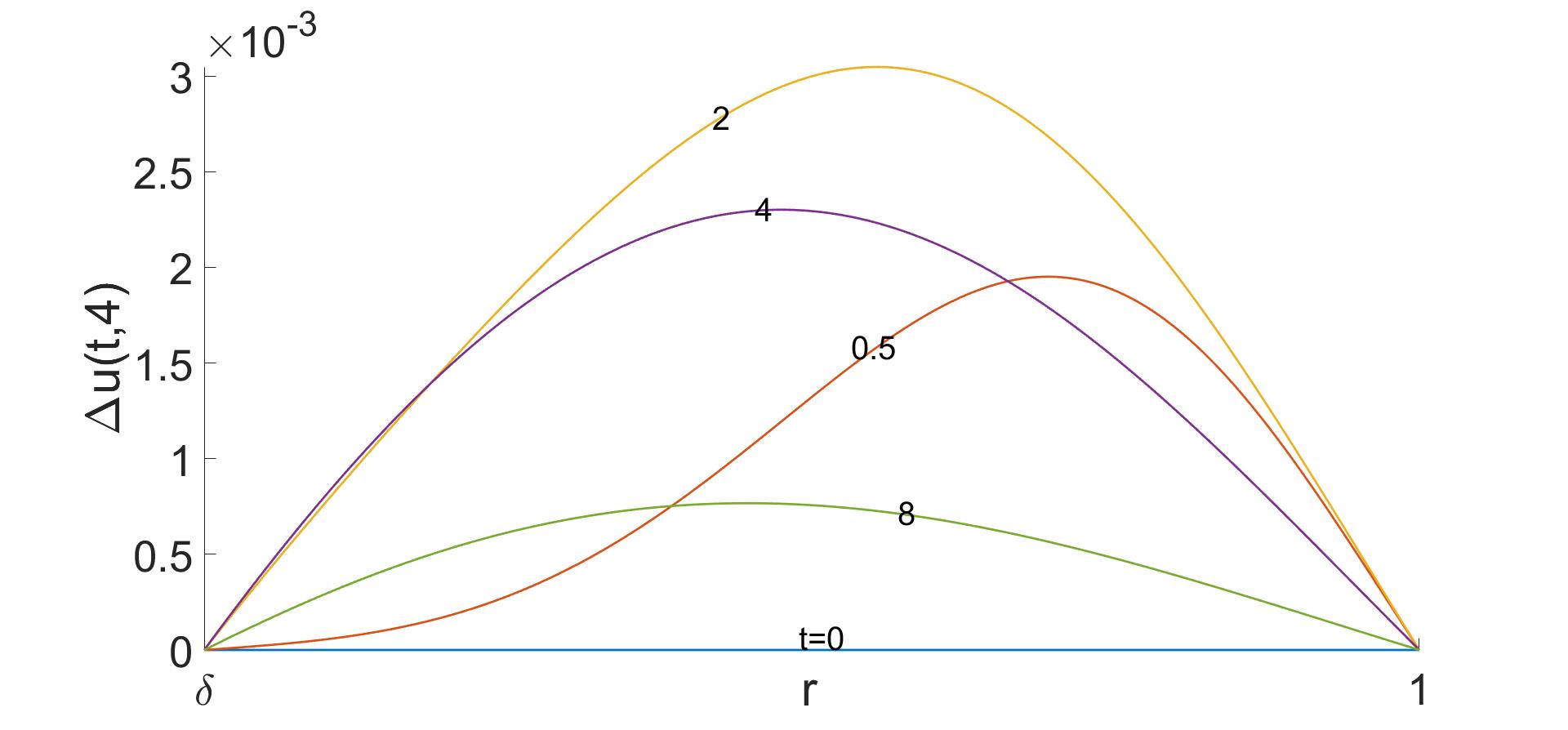}
\caption {Computed differences $\Delta b_z(r,t;\,q)$,  $\Delta rb_{\theta}(r,t; \,q)$, $\Delta \rho(r,t; \,q)$ and $\Delta u(r,t;\,q)$ between the case $q=4$ and the case $q=0$ for $\kappa=0.01$ at times $t=0,\,0.5,\,2,\,4,\,8$. The left column shows, for reference, the evolution of  $b_z(r,t)$, $rb_{\theta}(r,t)$, $\rho(r,t)$ and $u(r,t)$ when $q=0$.} 
\label{Fig_Deltas_eta0d01}
\end{figure}
\begin{figure}
\centering
%[width= \textwidth,  trim=0mm 140mm 0mm 0mm]
\includegraphics[scale=0.123]{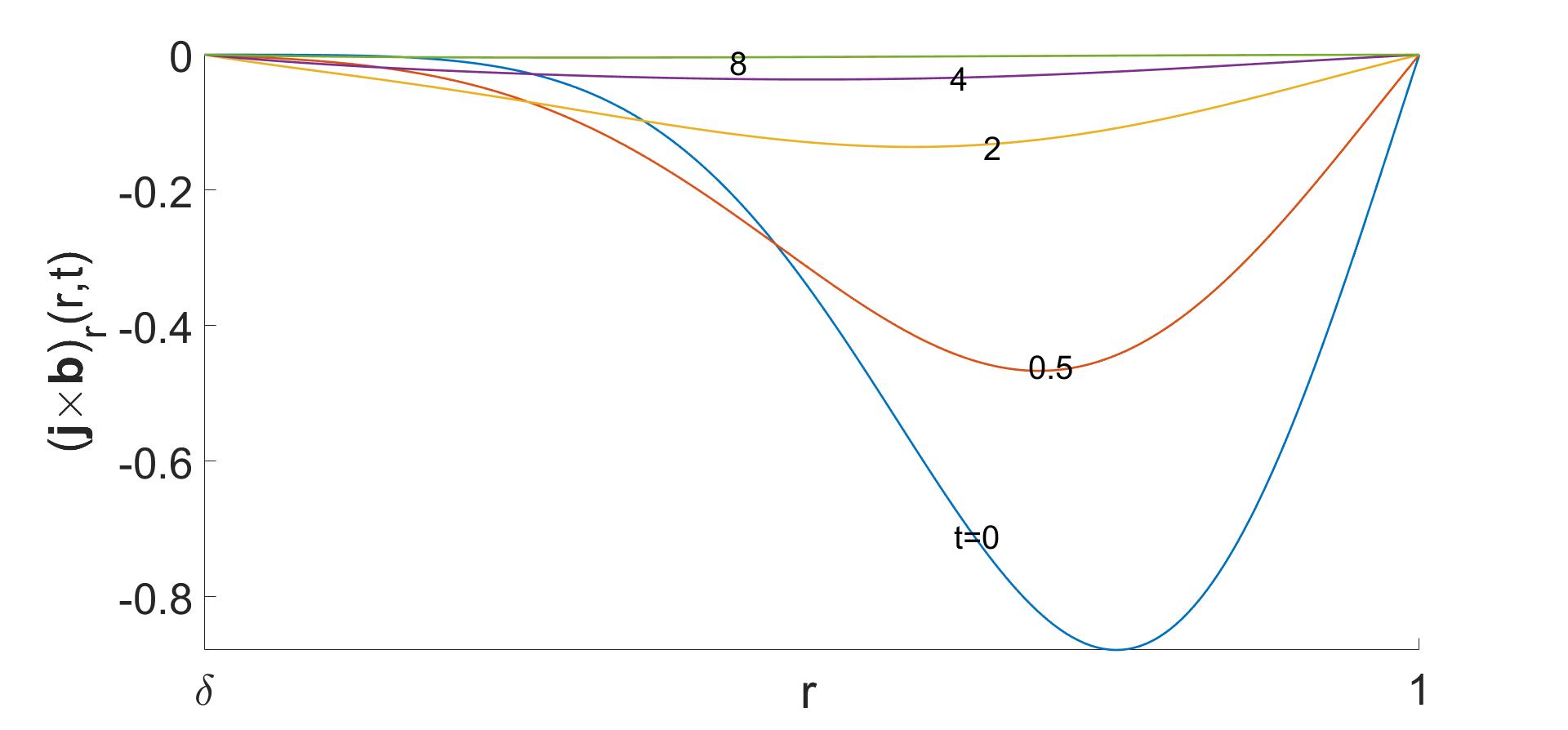}\includegraphics[scale=0.123]{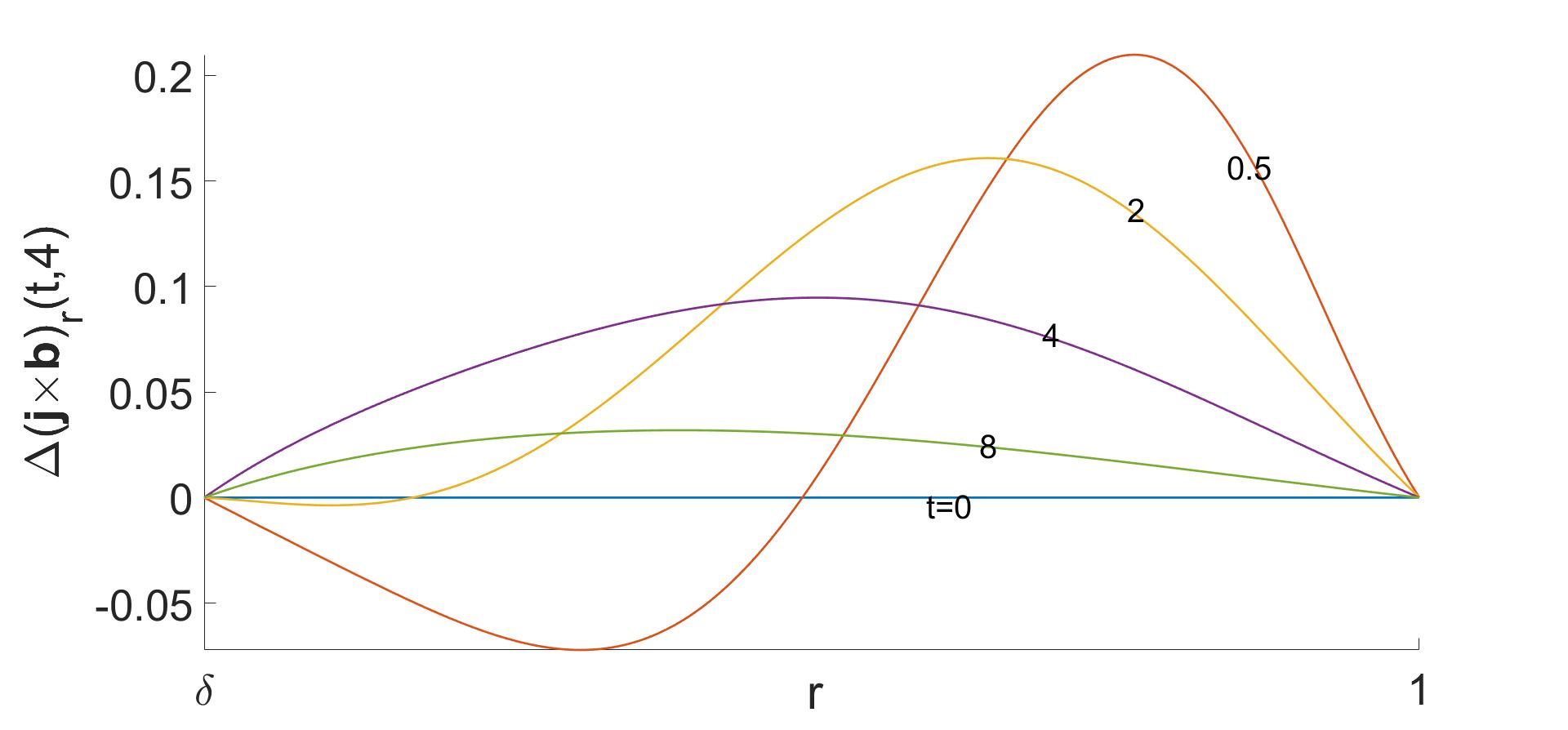}
\includegraphics[scale=0.123]{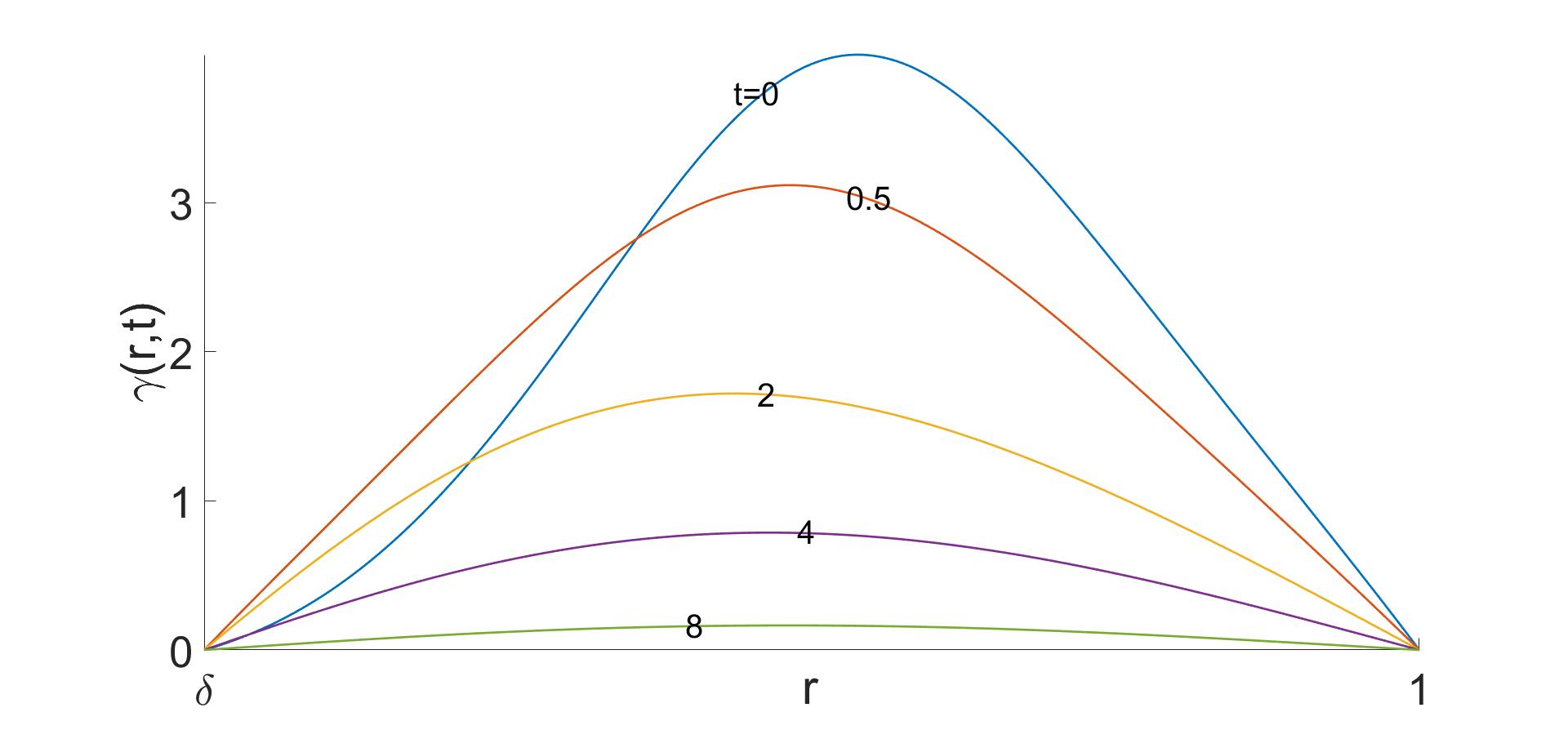}\includegraphics[scale=0.123]{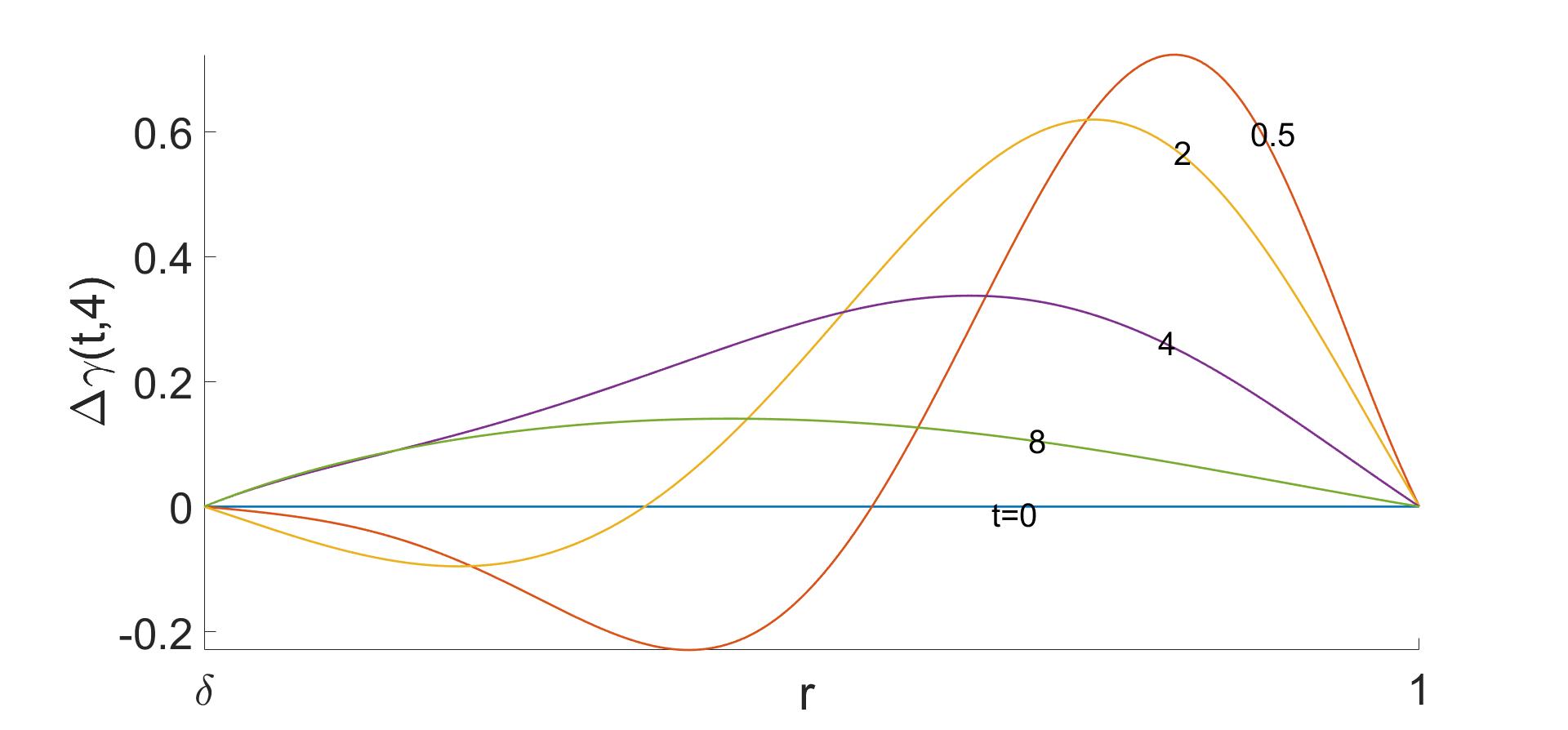}
\caption {Computed differences $\Delta (\mathbf{j}\times\mathbf{b})_{r}(r,t;\,q)$ and $\Delta \gamma(r,t;\,q)$ for same parameter values as in Fig.~\ref{Fig_Deltas_eta0d01}. The left column shows, for reference, the evolution of  $(\mathbf{j}\times\mathbf{b})_{r}(r,t)$ and $\gamma(r,t)$ when $q=0$.} 
\label{Fig_Deltas2_eta0d001}
\end{figure}
%

%%%%%%%%%%%%%%%%%%%%%%%%%%%%%%%%%%%%%%%%%%%%%%%%%%%%%%

 %

Numerical integration for $q\lesssim 3.5$ were quite regular, and not greatly different from the situation when $q=0$.  However, as might be expected from the above discussion,  numerical instabilities that are difficult to control appear when $q\gtrsim 4$.  This however is the regime that must be investigated in seeking possible reversal of $b_z(r,t)$.  

This led us to adopt a  controllable numerical procedure, specifically  4th-order finite-differences in the radial direction and  2nd-order Adams-Bashforth time-stepping with Crank-Nicolson treatment of the diffusive terms.  The results for $q=0$ were as expected in complete agreement with those obtained using \emph{Mathematica}.  We   focus  first on the near-critical situation when  $q=4$. 
Let $\rho(r,t;\,q)$ denote the density computed for any particular value of $q$, and let 
\begin{equation}
\Delta \rho(r,t; \,q)\!=\!\rho(r,t;\,q)\!-\!\rho(r,t;\,0);
\end{equation}
similarly for $\Delta b_{z}(r,t;\,q), \Delta r b_{\theta}(r,t;\,q), \Delta u(r,t;\,q),\,\Delta (\mathbf{j}\times\mathbf{b})_{r}(r,t;\,q)$ and $\Delta \gamma(r,t;\,q)$.
Figs.~4 and 5 show numerical results for $\kappa=0.01$; the left-hand columns show curves for $q=0$, while the right-hand columns show the differences $\Delta b_{z}(r,t; \,q)$, etc.
 
Note first from the last row of Fig.~4 that the negative (pinching) velocity is initially decreased in magnitude by the $\alpha$-effect when $q=4$; it actually becomes weakly positive near $r=1$ for $t\gtrsim 2$ and is positive  over the whole range $(\delta,1)$ for $t\gtrsim 2.5$. This implies a corresponding net decrease in the inward transport of mass; however, the decrease in transport for the magnetic field component $b_{z}(r,t)$ is more than compensated by the  direct action of the $\alpha$-effect. When $q=4$,  $b_z(r,t)$ decreases much more rapidly than when $q=0$ (by a factor of about 3) near $r=1$ in the early stage of relaxation;  $r b_{\theta}(r,t)$ increases more rapidly near $r=1$, but by a more modest amount ($\sim6\%$).      The function $\gamma (r,t)$ defined by eqn.~(\ref{gamma}) is also changed by  $\sim 25\%$ when $q=4$, but there is no apparent tendency for  $\gamma(r,t)$ to become more uniform.

\subsection{Results for $q=5.5$}
\begin{figure}
\centering
%[width= \textwidth,  trim=0mm 140mm 0mm 0mm]
\includegraphics[scale=0.2]{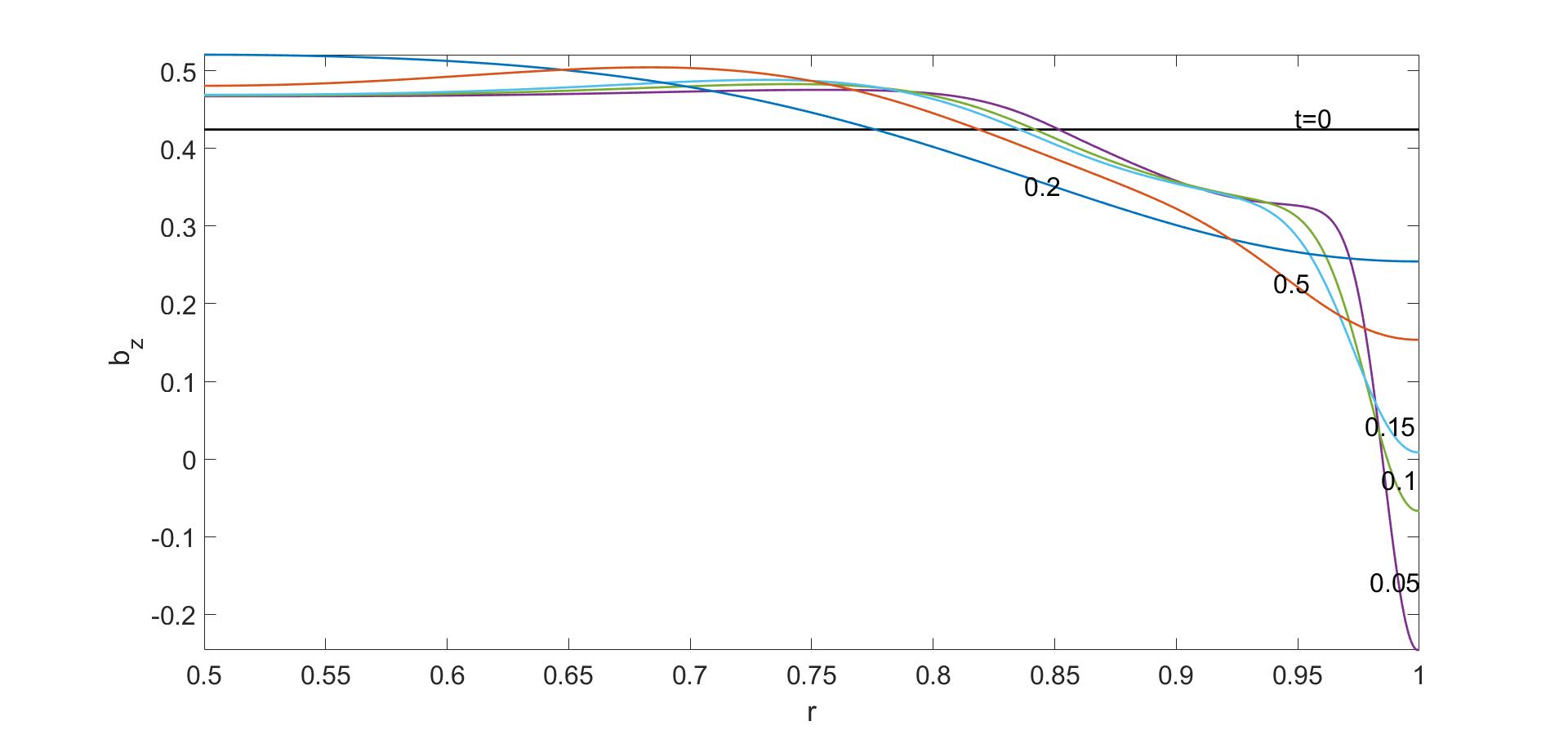}
\caption {Early stage of evolution of $b_z(r,t)$ for q=5.5, $\kappa=0.01$, and other parameter values  as in Fig.~\ref{Fig_Deltas_eta0d01}.}
\label{q_above_qc}
\end{figure}

As indicated above, we  anticipated numerical problems for $q\gtrsim 4$, and we did indeed run into these. Typically, in the range of $q$ between 4 and 6, a packet of oscillations in $b_z$ of very short wavelength appears at $t=0+$ in the region of negative diffusivity near $r=1$.  These oscillations move inwards, in tandem with $r b_{\theta}$ which decreases till the local diffusivity $\kappa(1-q b_{\theta}^2)$ becomes positive, at which stage the oscillations  in $b_z$ are damped out,  the subsequent evolution being quite smooth.   Fig.~\ref{q_above_qc} shows this subsequent evolution for $q=5.5$. 
Note that already at the early time $t=0.05$, $b_z$ is negative near $r=1$, presumably a consequence of the early negative diffusivity in this region.  The field then relaxes back, becoming positive at $r=1$ by time $t=0.15$.  

The short period of reversed $b_z$ near $r\!=\!1$ is interesting in the context of the reversed-field pinch. However, we can't be  certain that this is a genuine physical effect rather than just a consequence of adopting a unphysical model for the $\alpha$-effect yielding a period of negative diffusion.  The behaviour for $t=0+$ is evidently non-analytic;  an asymptotic treatment of the behaviour as $t\downarrow 0$ is presented in Appendix A.

\section{Conclusions}
We have investigated the relaxation of an axisymmetric magnetic field having both axial and toroidal components in a cylindrical geometry with perfectly conducting boundaries.  The density is assumed very small and fluid pressure is neglected compared with magnetic pressure.  A purely radial flow is driven by the Lorentz force and energy is dissipated by viscosity.  In the zero-resistivity limit, the field rapidly relaxes to a force-free state.  When weak resistivity is taken into account, the initial rapid relaxation is followed  by slow decay of the field which is constrained to remain  nearly force-free with ${\bf j}=\gamma(r,t){\bf B}$. However $\gamma(r,t)$ is quite strongly non-uniform, so  this is not a Taylor state.  

In \S 8, we have  explored the possibility that an $\alpha$-effect, with $\alpha$ proportional to ${\bf j}\cdot {\bf b}$, might be capable of causing axial field reversal near the outer boundary where the toroidal field component $b_{\theta}$ is initially strong.  We have found that if this $\alpha$-effect is sufficiently strong, it can produce a region of negative effective diffusivity of the $b_z$-field near the outer boundary; this can instantaneously generate high-frequency short-wavelength oscillations which are rapidly damped  as they move into the interior region of positive diffusivity.  A transitory reversal of $B_z$ occurs near the outer boundary during this process.

There are serious numerical difficulties in handling such a situation; nevertheless, this work points to one possibility whereby a reversed axial field, as observed in the reversed field pinch (Taylor 1974), can be dynamically generated from an initially uniform axial field, through the combined action of pinching by the $b_{\theta}$-field and a suitably contrived $\alpha$-effect.  Of course, it would be desirable to derive a correct form of this $\alpha$-effect, through investigation of the turbulence that results from instabilities of the relaxing field.  Work is ongoing on this aspect of the problem.

{\bf {Acknowledgments}}.
The partial funding of the Ministry of Science and Higher Education of Poland within the grant no IP 2014 031373 and statutory activities No 3841/E-41/S/2015 is gratefully acknowledged. This work was also partially funded from the Leading National Research Centre (KNOW) received by the Centre for Polar Studies in Poland for the period 2014-2018.

 %

%%%%%%%%%%%%%%%%%%%%%%%%%%%%%%%%%%%%%%%%%%%%%%%%%%%%%%%%%%%%%%%%%%%%%%%
%%%%%%%%%%%%%%%%%%%%%%%%%%%%%%%%%%%%%%%%%%%%%%%%%%%%%%%%%%%%%%%%%%%%%%%
%%%%%%%%%%%%%%%%%%%%%%%%%%%%%%%%%%%%%%%%%%%%%%%%%%%%%%%%%%%%%%%%%%%%%%%
%%%%%%%%%%%%%%%%%%%%%%%%%%%%%%%%%%%%%%%%%%%%%%%%%%%%%%%%%%%%%%%%%%%%%%%
%%%%%%%%%%%%%%%%%%%%%%%%%%%%%%%%%%%%%%%%%%%%%%%%%%%%%%%%%%%%%%%%%%%%%%%

%\begin{acknowledgments}

%\end{acknowledgments}

%\nocite{*}
%\bibliography{sorsamp}

% Produces the bibliography via BibTeX.

\section*{References}
\begin{harvard}
\item[]  Arnold, V.I. (1974) The asymptotic Hopf invariant and its applications. \emph {Proc. Summer School in Diff. Eqs.}  Erevan, Armenia [In Russian]. 229-256.
\item[]  Bajer, K. \& Moffatt, H. K. (2013) Magnetic relaxation, current sheets, and structure formation in an extremely tenuous fluid medium. \emph {Astrophys. J.} {\bf 779}, 169-182.
\item[] Bender, C.M. \& Orszag, S.A. (1978) \emph {Advanced Mathematical Methods for Scientists and Engineers.}  McGraw?Hill, New York.
\item[]  Bennett, W. H. (1934) Magnetically self-focussing streams. \emph {Phys. Rev.} 45.12: 890.
\item[] Furth, H. P., Killeen, J. \& Rosenbluth, M. N. (1963) Finite-resistive instabilities of a sheet pinch \emph{Phys. Fluids},  459-
\item[] Mizerski, K. (2017) Large scale EMF in current sheets induced by tearing modes, \emph{Fluid Dyn. Res.}, Submitted.
\item[] Moffatt, H. K. (1985)  Magnetostatic equilibria and analogous Euler flows of arbitrarily complex topology. Part 1. Fundamentals. \emph {J. Fluid Mech.} {\bf 159}  359-378.
\item[] Moffatt, H. K. (2015) Magnetic relaxation and the Taylor conjecture. \emph {J.Plasma Phys.} 81.06: 905810608.
\item[] Taylor, J. B. (1974) Relaxation of toroidal plasma and generation of reverse magnetic fields, \emph {Phys. Rev. Lett.} {\bf 33}, 1139-1141.
\end{harvard}

\appendix

\section{Short-time asymptotics}

The short-time behaviour of  $b_z(r,t)$ is bound to be non-analytic, since even the linear diffusion term introduces short-time dependence of the sort $\sim\exp(-r^{2}/4\kappa t)$. Although the problem is nonlinear, to get some insight into this short-time behaviour,  we suppose that the  controlling factor of $b_z$, i.e. its most rapidly changing component (Bender \& Orszag 1978), has the asymptotic WKB form
\begin{equation}\label{WKB_form_bz}
b_z(r,t) - b_z(r,0)\sim \mathrm{e}^{\mathcal{T}(t)\mathcal{R}(r)}\quad \textnormal{as}\,\,t\rightarrow 0,
\end{equation}
where $\mathcal{T}(t)$ and $\mathcal{R}(r)$ are functions to be determined. We further assume, that since the velocity is weak throughout the entire evolution, the terms involving $u(r,t)$ in (\ref{b_z_eqn_final}) and (\ref{b_theta_eqn_final}) can be neglected at the initial stage leaving only the  effect of diffusion in the $b_{\theta}$ component.  The behaviour at small $t$ of $b_{\theta}$ can therefore be expressed by a simple formula $b_{\theta}\approx b_{\theta}(r,0) + \beta(r,t)$ where $\beta(r,t)\rightarrow 0$ as $t\rightarrow0$. 
%\footnote{In fact the small $t$ expression for $rb_{theta}$ satisfying a diffusion equation is $rb_{\theta}(r,t\ll1) - rb_\theta(r,0) \sim \mathrm{e}^{-r^2/4\kappa t} + f(t)g(t)$, and since $ rb_\theta(r,0)$ is a non-trivial function of $r$  this expression involves additional terms of type $f(t)g(r)$, where $f(t)$ is a slowly varying function of $t$ and $g(r)$ involves derivatives of $b_{\theta}(r,0)$; however, the $t$-dependent terms in the small $t$ expansion of $b_{\theta}(r,t)$ do not affect our leading order analysis of the short-time behaviour of $b_z(r,t)$.}.
Substituting the asymptotic expressions  for the magnetic field components into equation (\ref{b_z_eqn_final}) we obtain, for small $t$,
\begin{equation}\label{WKB1}
\mathcal{R}\dot{\mathcal{T}} \approx \kappa K(r)\mathcal{R}^{\prime 2}\mathcal{T}^2,\quad \textnormal{where}\,\,K(r)=1-qb_{\theta}(r,0)^2,
\end{equation}
and where the dot denotes a time derivative and the prime a derivative with respect to $r$. It follows that,
\begin{equation}
\frac{\dot{\mathcal{T}}}{\mathcal{T}^2}=\mathrm{cst.}=\frac{\kappa K(r)\mathcal{R}^{\prime 2}}{\mathcal{R}}.\label{WKB2}
\end{equation}
The solution of the above equations is
\begin{equation}
\mathcal{T}(t)\mathcal{R}(r) = -\frac{1}{4\kappa t}\left\{ \int_\delta^r \left[K(r)\right]^{-1/2}\mathrm{d}r +\mathrm{cst.}\right\}^2.\label{WKB_solution}
\end{equation}

We have seen that increasing the value of $q$ leads to negative diffusion in the evolution of $b_z(r,t)$. As long as $q\lesssim1/b_{\theta}(1,0)^2=4$ the short-time asymptotics of $b_z(r,t)$  involve no significant irregularities. For $q$ above the critical value, the negative diffusion introduces serious irregularities and the point $r=r_c$ at which the diffusion coefficient $K(r)$ changes sign becomes a singular (critical) point. In the vicinity of $r_c$, the WKB solution ceases to be valid since gradients become infinite and a critical layer\footnote {With $\xi=(r-r_c)/\lambda$, $\lambda\ll1$ being the critical layer thickness, one can expand $1-qb_{\theta}(r_c+\xi\lambda,0)^2\approx-q\xi\lambda[b_{\theta}(r,0)^2]^{\prime}_{r=r_c}$ and the $b_z$ critical-layer equation is $\partial_{t^*} b_z\approx-\xi\partial^2_{\xi}b_z$ with $\lambda=q\kappa \tau[b_{\theta}(r,0)^2]^{\prime}_{r=r_c}$ and $t=t^*\tau$ -- note that as the time scale $\tau$ increases the critical layer thickens; this equation may be solved in terms of a Laplace transform in time with $r$-dependence of each Laplace mode $\sim\mathrm{e}^{\sigma \tau}$ in the form $\sim\xi^{1/2}J_1(\sigma\xi^{1/2})$, and matched to the WKB solutions on both sides, i.e for $r<r_c$ and $r>r_c$.}  of width $\mathcal{O}(\lambda)\ll 1$ is required to match the $b_z$-derivatives across $r_c$.  A continuous WKB solution for $t\ll1$ for the controlling factor, with a jump in the first derivative at $r=r_c$, satisfying the boundary conditions (\ref{bc2}), can  be found in the form
\begin{equation}
b_z(r,t) - b_z(r,0)\sim \exp\left\{-\frac{1}{4\kappa t}\left[ \int_\delta^r \frac{\mathrm{d}r}{\sqrt{K(r)}} \right]^2\right\},\label{sol_region1}
\end{equation}
for $r<r_c$, and
\begin{eqnarray}
b_z(r,t) - b_z(r,0)\sim & & \nonumber
\end{eqnarray}
\begin{equation}
\exp\left\{-\frac{1}{4\kappa t}\left[I_1^2-\mathcal{I}(r)^2+2I_2\mathcal{I}(r) \right]\right\}\cos\left[ \frac{\sqrt{I_1^2+I_2^2}}{2\kappa t}\left(\mathcal{I}(r)-I_2\right) \right],\label{sol_region2}
\end{equation}
for $r>r_c$, where 
\begin{equation}\label{KI}
 \mathcal{I}(r)= \int_{r_c}^{r} \frac{\mathrm{d}r}{\sqrt{-K(r)}},\quad I_1= \int_\delta^{r_c} \frac{\mathrm{d}r}{\sqrt{K(r)}},\quad I_2= \int_{r_c}^1 \frac{\mathrm{d}r}{\sqrt{-K(r)}}.
\end{equation}
(Strictly, the upper limit in $I_1$ should be $r_c-\lambda$, and the lower limit in $I_2$ and $\mathcal{I}(r)$ should be $r_c+\lambda$, due to the presence of the critical layer at $r_c$, which matches the derivatives of the WKB solutions on either side of the critical point $r_c$, but this correction is negligibly small.)
\begin{figure}
\centering
%[width= \textwidth,  trim=0mm 140mm 0mm 0mm]
%\includegraphics[scale=0.123]{WKB_CF_t0d01_1.jpg}\includegraphics[scale=0.123]{WKB_CF_t0d06_1.jpg}
\includegraphics[scale=0.123]{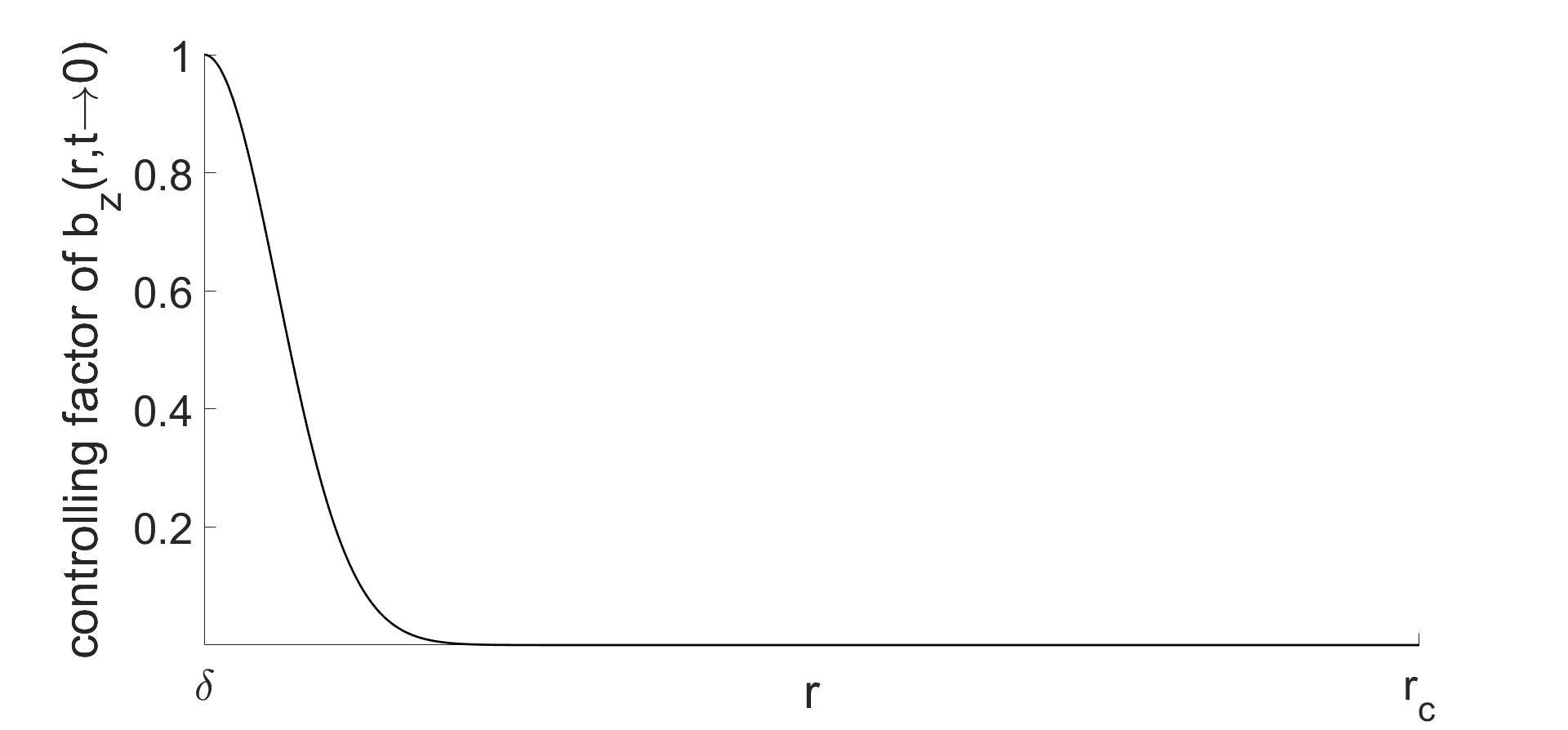}\includegraphics[scale=0.123]{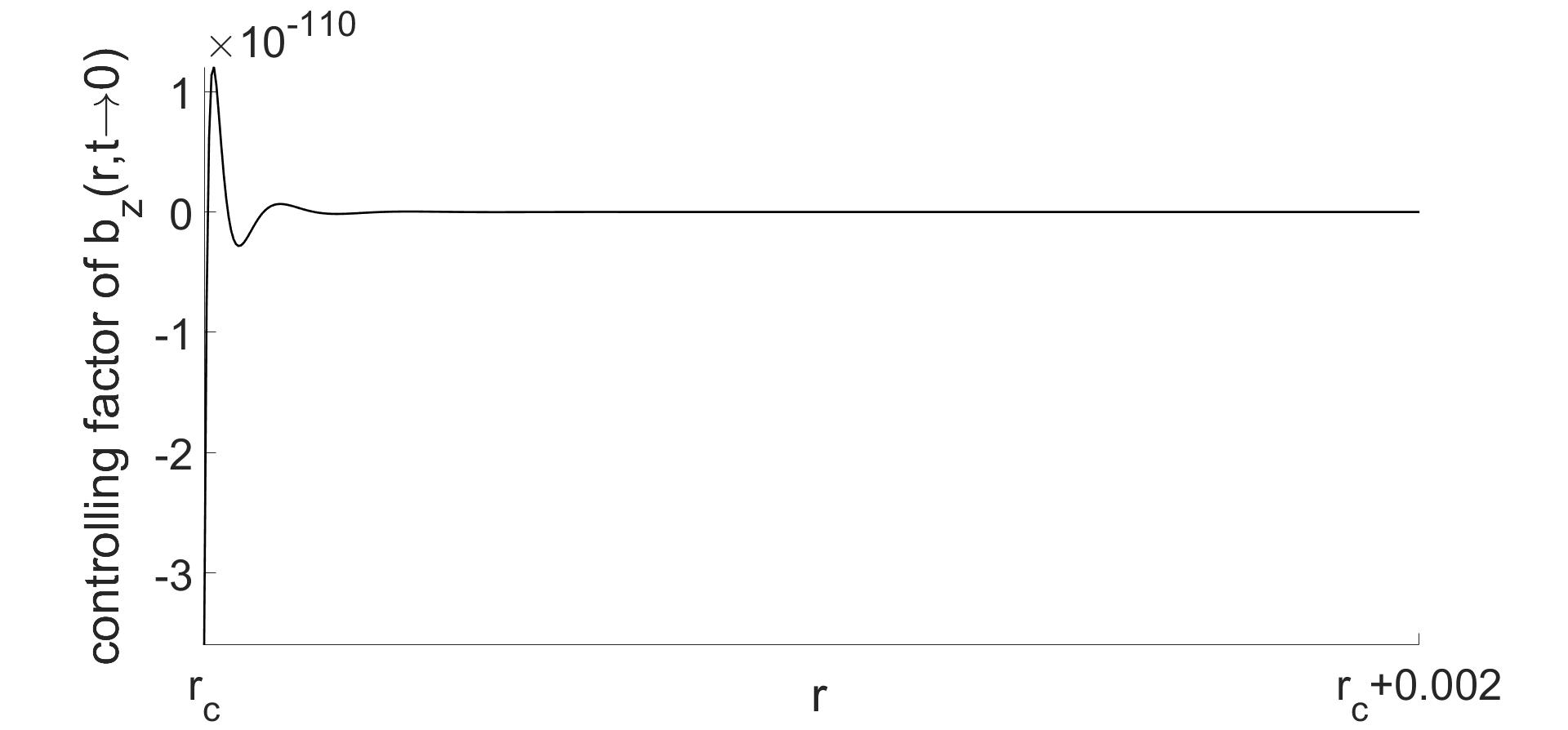}
\includegraphics[scale=0.123]{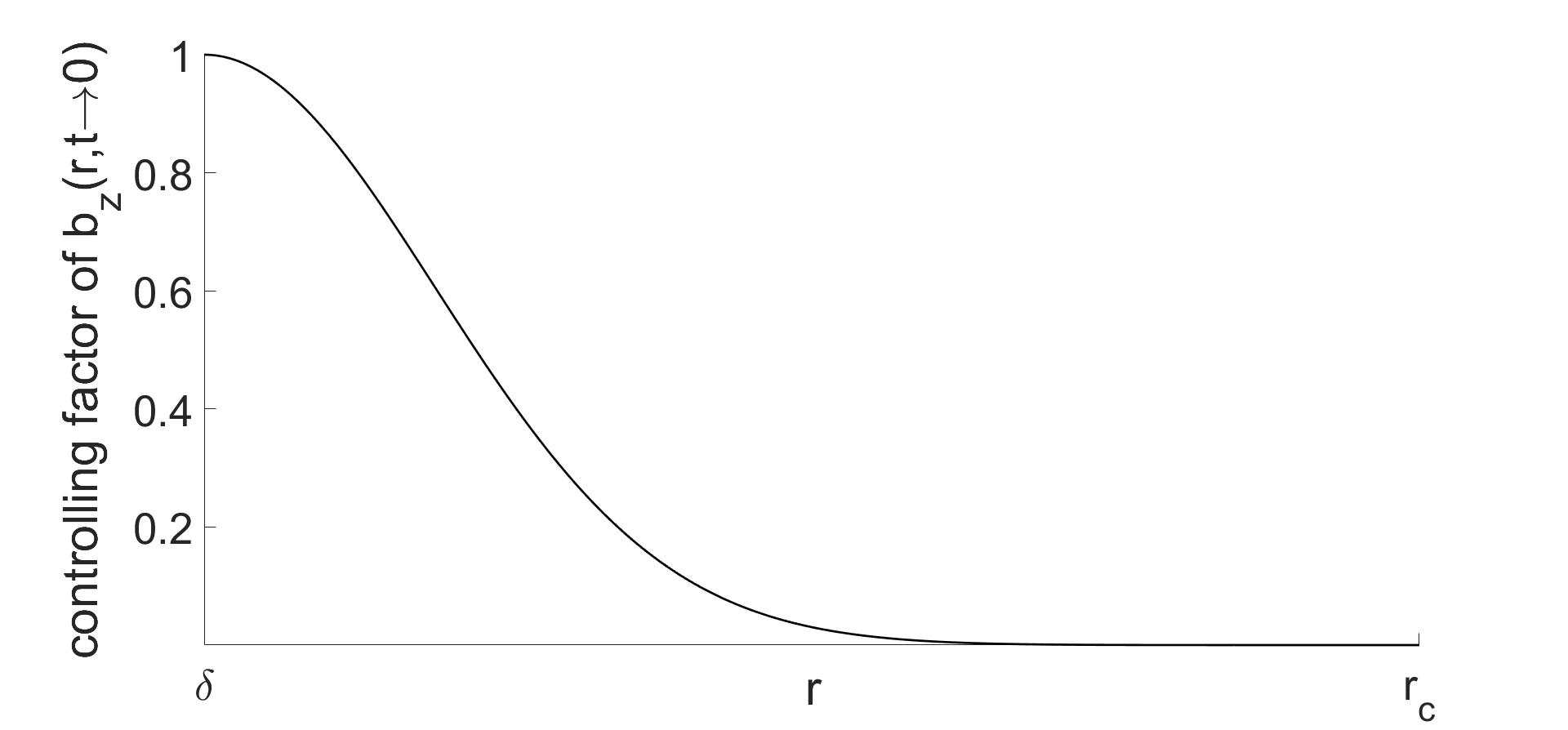}\includegraphics[scale=0.123]{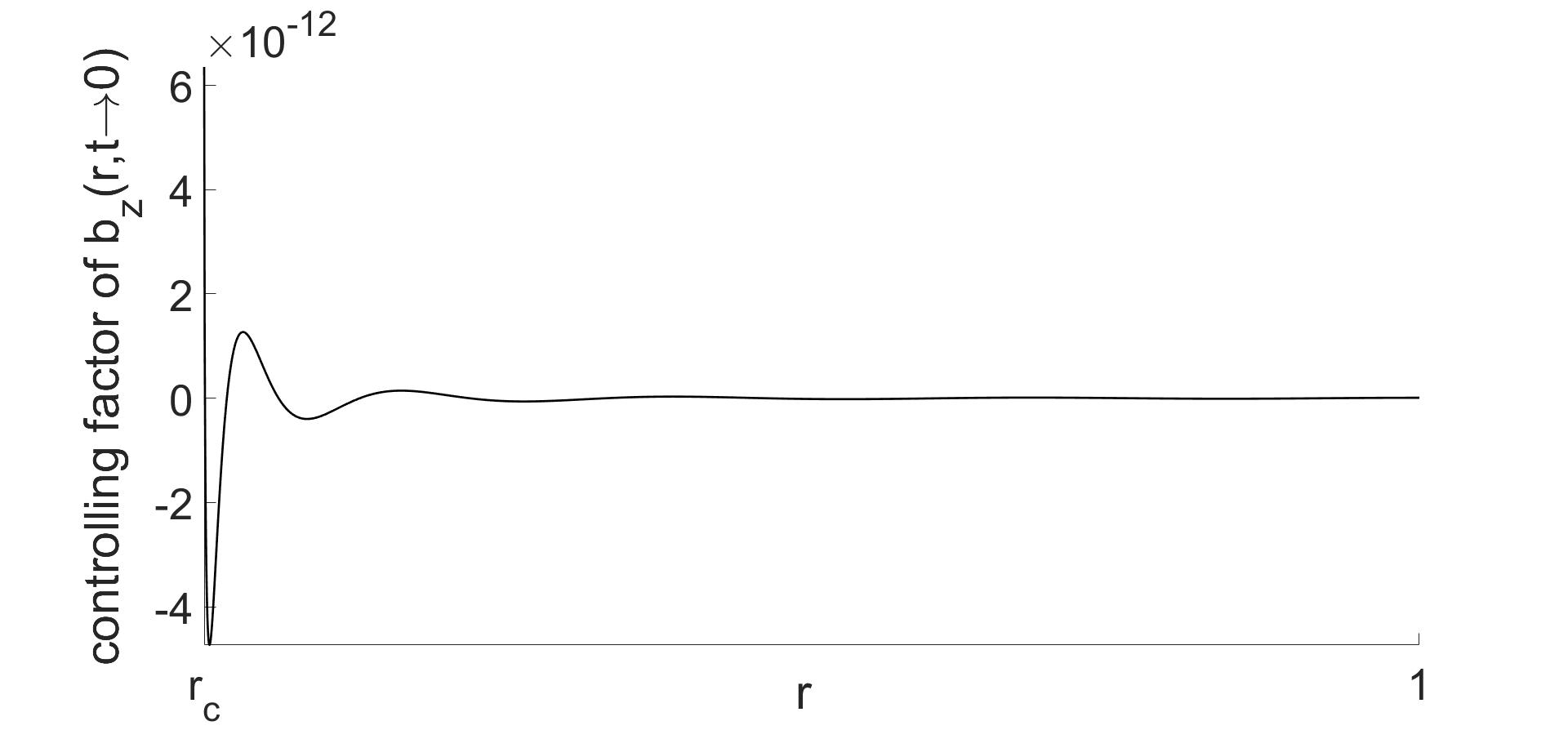}
\caption {Plots of the asymptotic $r$-dependence of the controlling factor, i.e. the most rapidly changing component of $b_z(r,t\ll1)$ at times $4\kappa t=0.001$ - top row, and $4\kappa t=0.01$ - bottom row, for $q=6$ $(r_c\approx0.8725)$ and other parameter values as in the numerical simulations (Fig.~\ref{Fig_Deltas_eta0d01}). The regions $r<r_c$ and $r>r_c$ are plotted separately in the left and right columns respectively. Note that in reality the sharp change in derivative at $r=r_c$ is smoothed out by a boundary layer, thickening with time.}
\label{WKB_solutions}
\end{figure}
It is clear, therefore, that for $r<r_c$ the solution is regular and controlled mainly by diffusion. However, for $r>r_c$  oscillations of very short $\mathcal{O}(t)$ wavelength appear, which is smallest near $r_c$ and increases with $r$;  since $I_1>I_2>\mathcal{I}(r)$ for any $q$, the amplitude of these oscillations  is exponentially small and decreases with increasing $r$.  They appear instantaneously at $t=0+$ and their wavelength increases with increasing $t$. At very short times they are strongly damped by the very small exponential term in (\ref{sol_region2}). The situation is depicted (for $q=6$) in Fig.~\ref{WKB_solutions}. 

 These fast  short-time oscillations  are just as described in \S 8.3 for the run with $q=5.5$. At the earliest stage of evolution we observed a small drop in the axial flux $\Phi_z$ of $b_z$ from $1$ to about $0.94$, which subsequently remained  constant; the toroidal flux $\Phi_{\theta}$ and the mass $\mathbb{M}$  experienced much smaller jumps at the same moment as that of $\Phi_z$, but then also remained constant. These jumps are an indication of unavoidable numerical inaccuracy at this earliest stage when the  extremely short wavelength oscillations cannot be adequately resolved by  numerical procedure, however much refined.%

\end{document}